\documentclass[prd,aps,a4paper,superscriptaddress,twocolumn,nofootinbib]{revtex4-2}
\usepackage{subfig}
\usepackage{graphicx}
\usepackage{color}
\usepackage{dcolumn}
\usepackage{bm}
\usepackage{slashed}
\usepackage{amsmath}
\usepackage{latexsym}
\usepackage{amssymb}
\usepackage{mathrsfs}
\usepackage{amsfonts}
\usepackage{url}
\allowdisplaybreaks
\begin{document}
\title{The Axial Gravitational Ringing of a Spherically Symmetric Black Hole Surrounded by Dark Matter Spike}

\author{Yuqian Zhao}
\affiliation{Institute for Frontiers in Astronomy and Astrophysics, Beijing Normal University, Beijing 102206, China}
\affiliation{Department of Astronomy, Beijing Normal University, Beijing 100875, China}
\author{Bing Sun}
\affiliation{%
CAS Key Laboratory of Theoretical Physics, Institute of Theoretical Physics, Chinese Academy of Sciences, Beijing 100190, China
}%
\author{Kai Lin}
\affiliation{%
Hubei Subsurface Multi-scale Imaging Key Laboratory, School of Geophysics and Geomatics, China University of Geosciences, Wuhan 430074, Hubei, China
}%
\author{Zhoujian Cao
\footnote{corresponding author}} \email[Zhoujian Cao: ]{zjcao@amt.ac.cn}
\affiliation{Institute for Frontiers in Astronomy and Astrophysics, Beijing Normal University, Beijing 102206, China}
\affiliation{Department of Astronomy, Beijing Normal University, Beijing 100875, China}
\affiliation{School of Fundamental Physics and Mathematical Sciences, Hangzhou Institute for Advanced Study, UCAS, Hangzhou 310024, China}

\begin{abstract}
Supermassive black holes at the center of each galaxy may be surrounded by dark matter. Such dark matter admits a spike structure and vanishes at a certain distance from the black hole. This dark matter will impact the spacetime near the black hole and the related ringing gravitational waves can show distinguished features of the black hole without dark matter. In the present work, we focus on the quasi-normal modes of the axial gravitational perturbation who dominate the ringdown process of the perturbed black holes surrounded by dark matter spikes. The relativistic modification results in less impact on QNMs. And the relative ringing frequency difference between the black holes with and without dark matter can be as large as $10^{-2}$. These features can be used in future gravitational wave detection about extremal mass ratio inspiral systems to probe the existence of dark matter around supermassive black holes.
\end{abstract}

\maketitle

\section{Introduction}
Black holes (BHs) are important predictions of general relativity (GR)~\cite{einstein1915feldgleichungen}. In the past century, many scientists have endeavored to explore the existing evidences and characteristics of black holes~\cite{Barack_2019}. Since the first detection of gravitational wave (GW) event in 2015~\cite{PhysRevLett.116.061102}, more than 90 directly observed black hole binary events have been detected~\cite{GWOSC,GWTC1,GWTC2,GWTC2.1,GWTC3}. The observations of the shadow from M87 and our galaxy also provided additional evidence for supermassive black holes (SMBHs)~\cite{1Akiyama_2019,2Akiyama_2019,3Akiyama_2019,4Akiyama_2019,5Akiyama_2019,6Akiyama_2019,7Akiyama_2021,8Akiyama_2021}.

The event horizon is the distinguishable feature of BHs. Event horizon, as a one-way causal boundary, prevents us from detecting the inside of a black hole~\cite{1916skpa.conf..424S}. Black holes in our universe may be affected by astronomical environments, such as the dark matter (DM) near the BHs~\cite{PhysRevD.96.083014,Nampalliwar_2021,Xu_2021}. It is believed that 90\% of the host galaxies of SMBHs are composed of DM~\cite{jusufi2020shadows}. It is interesting to ask how the DM around the BHs impacts the spacetime and GWs related to these BHs.

The exploration of DM has a long history. The numerical results from $N$-body cosmological simulations suggest the density distribution of DM is peaked near the center of galaxies and decreases as a power of $1/r$ with $r$ the distance from the halo's center, as given by~\cite{PhysRevD.106.044027}:
\begin{equation}\label{rho}
\rho(r)=\rho_0(r / r_0)^{-\gamma_0}\left(1+(r / r_0)^{\alpha_0}\right)^{(\gamma_0-\beta_0) / \alpha_0}
\end{equation}
where $r_0$ and $\rho_0$ are the scale factors determined by the numerical fitting such as~\cite{10.1093/mnras/stz1698}, while $\alpha_0$, $\beta_0$ and $\gamma_0$ are the parameters determined by the choice of models with $(\alpha_0,\beta_0,\gamma_0)=(1,4,1)$ for Hernquist profile and $(\alpha_0,\beta_0,\gamma_0)=(1,3,1)$ for Navarro-Frenk-White (NFW) profile~\cite{hernquist1990analytical,Navarro_1997}.

With the presence of a BH, the density distribution of DM  changes. In an early pioneering work~\cite{PhysRevLett.83.1719}, a Newtonian method is used to calculate the redistribution of cold DM near the center of galaxies. The BH accretion makes the DM form a cusp (``spike") structure. For spherically symmetric BHs, the density peaks near $r\sim 4 R_s$ with $R_s$ the Schwarzschild radius and has a steep cutoff at $r=4 R_s$, below which the density of DM vanishes due to annihilation or dropping into the BHs. With the relativistic modification~\cite{PhysRevD.88.063522}, the density profile has similar characteristics, but the cutoff radius will be changed to $r=2R_s$. 

In~\cite{PhysRevD.106.044027}, both Hernquist and NFW models with or without relativistic modifications are introduced to explore the impacts of the DM spike on the extreme mass-ratio inspirals (EMRIs) GW waveform. And the relativistic modifications are found to have a positive impact on the detectability of DM for both models. In the current work, we use the same profile to explore the impacts of the ringdown waveform, specifically, the quasi-normal modes (QNMs) of the perturbed BHs~\cite{chandrasekhar1985mathematical,kokkotas1999quasi,Hans-PeterNollert_1999,Berti_2009,RevModPhys.83.793}. This DM profile takes the BHs Eddington accretion effect into consideration instead of Bondi accretion effect \cite{2022JCAP...08..032F}. Not like Refs.~~\cite{PhysRevD.89.023506,deluca2023superfluid} this DM profile does not consider the self-interactions of DM. Different DM profile may change the QNMs properties. 

The detection of ringdown signals with ground-based GW detectors is not satisfactory due to the low signal-to-noise (SNR)~\cite{PhysRevLett.129.111102}. Taking the advantage of high SNR, the impacts of DM spike on ringdown of SMBHs may be detected by the future space-based GW detectors including LISA, Taiji, Tianqin and DECIGO~\cite{ruan2020taiji,PhysRevD.100.044042,Moore_2015}. As an example, we will discuss the detectability of such signals for the Sgr$\text{A}^*$ BH at the center of Milky Way~\cite{10.1093/mnras/stz1698}, whose ringdown frequency exactly falls into the sensitive frequency bands of space-based GW detectors~\cite{PhysRevD.73.064030}.

In~\cite{PhysRevD.104.104042,ZHANG2022101078,PhysRevD.104.124082,PhysRevD.105.L061501,KONOPLYA2021136734}, the impacts of the cold DM halo on the ringdown process are explored. And the impacts of the DM spike with initial profile Eq.~\eqref{rho} and $\beta_0=\gamma_0$ on the QNMs of scalar perturbations can be found in~\cite{Daghigh_2022}. Here we extend the investigation to axial gravitational perturbations with Hernquist profile $(\alpha_0,\beta_0,\gamma_0)=(1,4,1)$ and Navarro-Frenk-White (NFW) profile $(\alpha_0,\beta_0,\gamma_0)=(1,3,1)$. And the impacts of different models and relativistic modification are also taken into consideration.

The structure of this paper is as follows: based on the profile of DM spike and using the TOV equations, we derive the modified gauge metric of spherically symmetric BHs in Sec.~\ref{sec:BHinDM}. We then derive the QNMs equation for axial gravitational perturbations in Sec.~\ref{sec:QNMeqn}. After that, we solve the QNMs equations with numerical methods and present the results in Sec.~\ref{sec:result}. We will use the Sgr$\text{A}^{*}$ with the DM spike as an example to explore the impacts of different DM spikes on QNMs and discuss the detectability of such impacts. Throughout the paper, we use the geometric unit system so that $c=G=1$, where $c$ is the speed of light and $G$ is the gravitational constant.

\section{Black holes surrounded by Dark Matter Spike}\label{sec:BHinDM}
\begin{figure}
    \includegraphics[width=\linewidth]{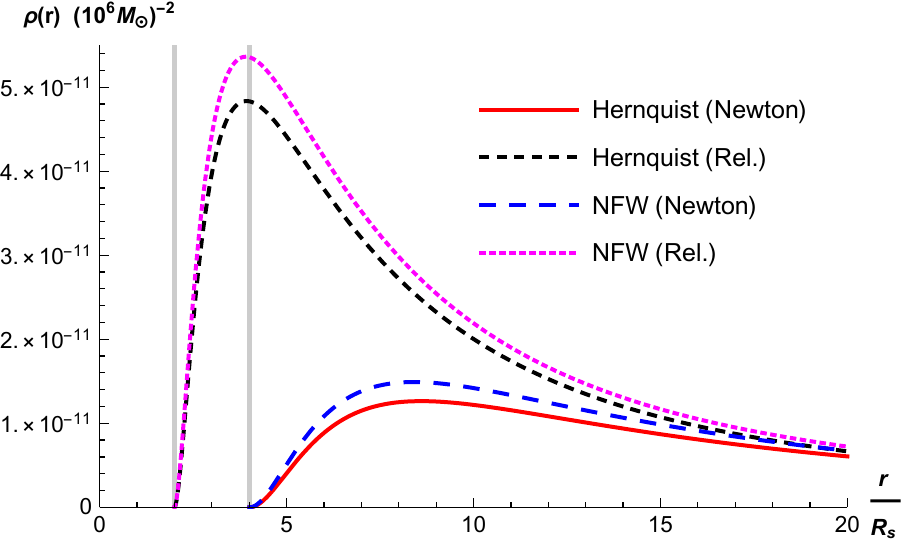}
    \caption{The DM profiles for Hernquist Newtonian model, Hernquist relativistic model, NFW Newtonian model and NFW relativistic model. The parameters are $M=10^6 M_\odot$, $\rho_0=0.3\text{GeV}/\text{cm}^3$, $r_0=20\text{kpc}$. The two vertical lines correspond $r=2 R_s$ and $r = 4 R_s$, below which the density vanishes for relativity and Newtonian DM respectively due to annihilation or dropping into the BH.}
    \label{fig:profile}
\end{figure}

In this section, we consider the spherically symmetric BHs surrounded by the DM spike and derive the modified metric. The most well-known solution for the spherically symmetric static BHs spacetime is the Schwarzschild BHs, which characterize the vacuum scenario. In general, such an isolated BH does not exist, particularly for SMBHs in the center of galaxies, which are more or less surrounded by some matters with complex distribution, such as DM. Then, a Schwarzschild BH in DM spike can be well described by the most general metric:
    \begin{equation}\label{metric}
        ds^2 = -f(r)dt^2 + \frac{dr^2}{g(r)} + r^2 (d \theta^2 + \sin^2 \theta d\varphi^2) \, .
    \end{equation}
and the contribution of the DM can be thought of as the energy-momentum tensor:
    \begin{equation}\label{Tmunu}
        T^\mu_\nu=\mathrm{diag}\left\{-\rho(r),p(r),p(r),p(r)\right\}
    \end{equation}
where $\rho(r)$ depends on the density distribution of DM. Based on~\cite{PhysRevD.106.044027}, we introduce the density profile of the DM spike as:
    \begin{equation}\label{profile}
        \rho(r)=\rho_R \left(1-\frac{4 \eta M}{r}\right)^w \left(\frac{\sigma M}{r}\right)^q
    \end{equation}
with $\sigma = 4.17\times 10^{11}$ and $\rho_R$ is the effective density of the DM spike given by:
    \begin{equation}\label{rhoR}
        \rho_R=A \times 10^\delta\left(\frac{\rho_0}{0.3 \mathrm{GeV} / \mathrm{cm}^3}\right)^\alpha\left(\frac{M}{10^6 M_{\odot}}\right)^\beta\left(\frac{r_0}{20 \mathrm{kpc}}\right)^\gamma
    \end{equation}
where the parameter $\eta=1(2)$ for relativistic (Newtonian) DM profiles and limits the location of a steep cutoff at $r=4 \eta M$ (corresponding to $2R_s$ and $4R_s$ respectively with $R_s$ the Schwarzschild radius), below which the DM density vanishes due to annihilation or dropping into the BH as shown in Fig.~\ref{fig:profile}. The parameters $(\alpha,\beta,\gamma,\delta)$ depend on whether the Hernquist model or the NFW model is chosen. The parameters $(A,w,q)$ are the fitting factors. $\rho_0$ and $r_0$ are the scale factors in Eq.\eqref{rho} and we choose the range the same as~\cite{PhysRevD.106.044027}, while $M$ is the mass of the BHs and the range is $10^{5} \sim 10^{9} M_{\odot}$ for SMBHs~\cite{RevModPhys.83.793}. Table~\ref{Tab:parameter} lists all the parameter values with the range of $\rho_R$.

\begin{table*}
    \caption{The parameters for DM density spike profiles. We choose the same values as~\cite{PhysRevD.106.044027} except the mass of BHs and we also calculate the realistic range of $\rho_R$. And we use ``Rel." to denote the models with relativistic modification.}
    \begin{ruledtabular}
    \begin{tabular}{c|cc|cc}\label{Tab:parameter}
    &Hernquist& Hernquist & NFW& NFW\\
    &(Newton)&(Rel.)&(Newton)&(Rel.)\\
    \hline \hline
     $\eta$ & $2$ &$1$&$2$&$1$\\
    $\alpha$ & $0.335$ & $0.335$ & $0.331$ &$0.331$\\
    $\beta$ & $-1.67$ & $-1.67$ & $-1.66$ & $-1.66$\\
    $\gamma$ & $0.31$ & $0.31$ & $0.32$ & $0.32$\\
    $\delta$ & $-0.025$ & $-0.025$ & $-0.000282$ & $-0.000282$\\
    $A(10^{-43} M_\odot^{-2})$ & $4.87$ & $7.90$ & $1.60$ & $6.42$\\
    $w$ & $2.22$ & $1.83$ & $2.18$ & $1.82$ \\
     $q$ & $1.93$ & $1.90$ & $1.98$ & $1.91$  \\
    $\rho_0\;(\text{GeV}/\text{cm}^3)$ & $0.1 \sim 0.5$ & $0.1 \sim 0.5$ & $0.1 \sim 0.5$ & $0.1 \sim 0.5$ \\
    $M\;( M_\odot)$ & $10^5\sim 10^9$  & $10^5\sim 10^9$  & $10^5\sim 10^9$  & $10^5\sim 10^9$ \\
    $r_0\;(\text{kpc})$ & $0.005\sim 50$ & $0.005\sim 50$ & $0.005\sim 50$ & $0.005\sim 50$\\
    $\rho_R M^2 $ & $1.14\times 10^{-32}$ & $1.85\times10^{-32}$ & $3.57\times10^{-33}$ & $1.43\times10^{-32}$\\
    & $\sim 7.08\times10^{-30}$ & $\sim 1.15\times10^{-29}$ & $\sim 2.66\times10^{-30}$ & $\sim 1.07\times10^{-29}$
    \end{tabular}
    \end{ruledtabular}
\end{table*}

\begin{figure*}
            \includegraphics[width=0.45\linewidth]{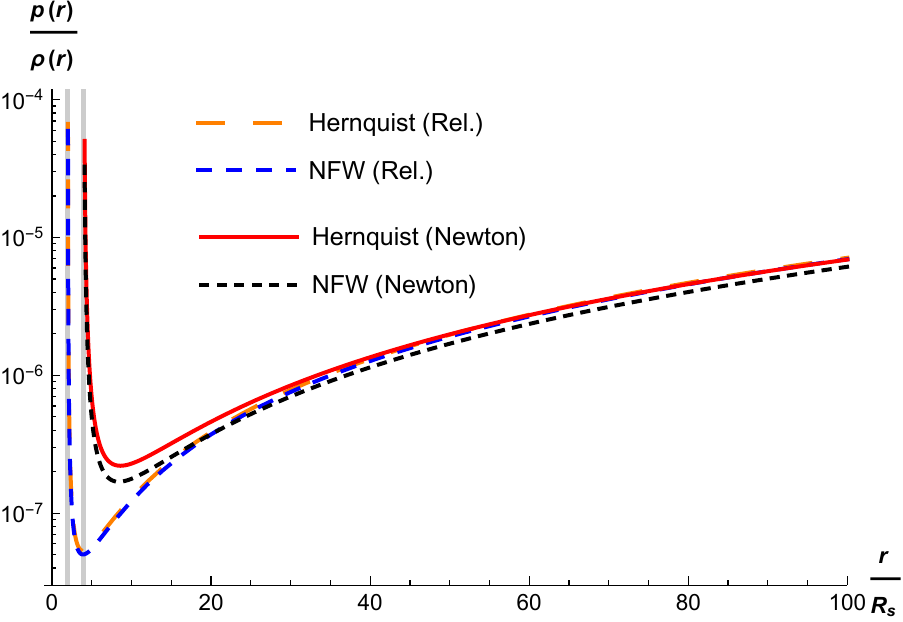}
        \hfill
    \subfloat{
            \includegraphics[width=0.45\linewidth]{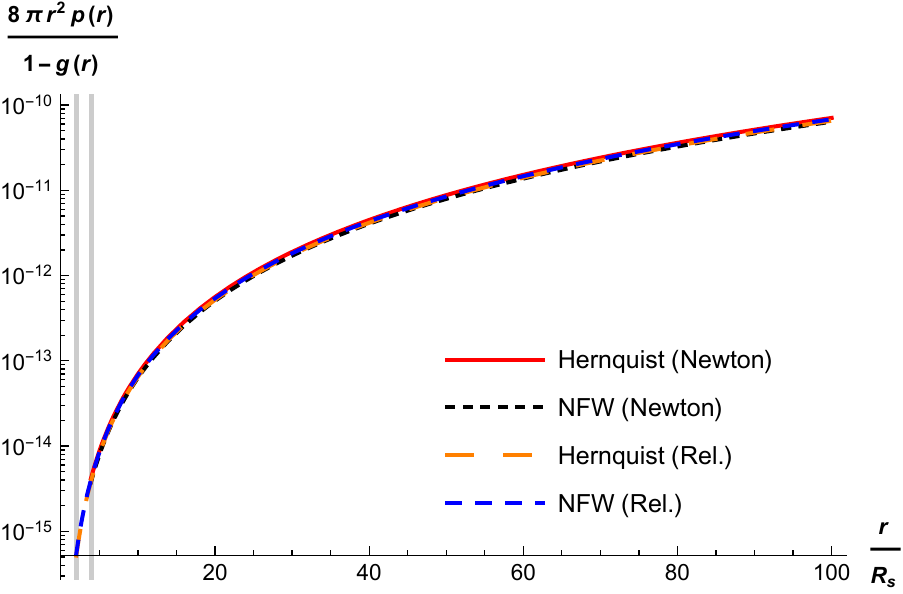}
        }
    \caption{Self consistence check of the assumption that the pressure $p(r)$ is ignorable in the spike region. The parameters $M=10^6 M_\odot$, $\rho_0=0.3\text{GeV}/\text{cm}^3$, $r_0=20\text{kpc}$ are used here to numerically solve Eq.~\eqref{TOV3}. The assumption that $p(r) \ll \rho(r)$ and $8 \pi r^2 p(r) \ll 1-g(r)$ is valid in this region as shown in the figure. The two vertical lines correspond $r=2 R_s$ and $r = 4 R_s$, blow which the density vanishes for relativity and Newtonian DM profile respectively.}
    \label{fig:pressure}
\end{figure*}

Inspired by the work of~\cite{Daghigh_2022}, we make the same assumption that the DM distributes like dust and the pressure $p(r)$ can be neglected with respect to $\rho(r)$ and $(1-g(r))/8\pi r^2$ as shown in Fig.~\ref{fig:pressure}. Einstein equations $G_{\mu\nu} =  8 \pi T_{\mu\nu}$ or the TOV equations~\cite{Daghigh_2022,Konoplya_2022,carroll2019spacetime} now read:
    \begin{align}\label{TOV1}
        r g^{\prime}(r) + g(r) - 1 + 8 \pi \rho(r) r^2 =0,\\
        \label{TOV2}
        f(r) g(r)-f(r)+r g(r) f^{\prime}(r) = 0,\\
        \label{TOV3}
        p^{\prime}(r)= - \rho (r)\frac{1-g(r)}{2r g(r)}.
    \end{align}
    By using Eq.~\eqref{profile}, we then integrate Eq.~\eqref{TOV1} from $4\eta M$ to the spatial infinity and obtain:
    \begin{align}
        &g(r)= 1- \frac{2M}{r}+ \frac{8\pi\rho_R (M \sigma)^q}{q-3}r^{2-q}\nonumber\\
        & \times {}_2 F_1 \left(q-3,-w,q-2m, \frac{4\eta M}{r}\right)\label{gsol}
    \end{align}
where ${}_2 F_1$ is the hypergeometric function~\cite{abramowitz3handbook,olver2010nist}. And we have chosen the constant of integration so that $g(4 \eta M)=1-\frac{2M}{r}$.

By substituting Eq.~\eqref{gsol} and Eq.~\eqref{profile} into Eq.~\eqref{TOV2}, $f(r)$ can be obtained as:
    \begin{align}
    &f(r)= \left(1-\frac{2 M}{r}\right) \times \exp \left\{\frac{1}{(q-3) \Gamma(q-3)}\right. \nonumber\\
    &\sum_{n=0}^{N_{\text{max}}}\left\{ 8 \pi \Gamma(q-2)\left(\rho_R r^2\right)\left(\frac{2 M}{r}\right)^n\left(\frac{M \sigma}{r}\right)^q \right. \nonumber\\
    & \left[\Gamma(q-3){ }_2 \tilde{F}_1\left(q-3,-w ; q-2 ; \frac{4 M \eta}{r}\right)\right.\nonumber\\
    &\left. -\Gamma(n+q-2)_2 \tilde{F}_1\left(n+q-2,-w ; n+q-1 ; \frac{4 M \eta}{r}\right)\right] \nonumber\\
    & -\pi\left(\rho_R M^2\right) 2^{7-n-2 q} \eta^{2-n-q} \sigma^q \Gamma(q-2) \Gamma(w+1)\nonumber\\
    &\left.\left.\left(\frac{\Gamma(q-3)}{\Gamma(q+w-2)}-\frac{\Gamma(n+q-2)}{\Gamma(n+q+w-1)}\right)\right\}\right\}\label{fsol}
    \end{align}
where $\Gamma$ is the Gamma function and $\, _2\tilde{F}_1\left(a,b;c;d\right)\equiv\, _2F_1\left(a,b,c,d\right)/ \Gamma(c)$ is the regularized hypergeometric function. Besides choosing suitable integration constant to satisfy $f(4\eta M)=1-\frac{2M}{r}$,
we use two tricks to simplify the integration process: (1) Noticing that $\rho_R M^2 \ll 1$ as shown in Table~\ref{Tab:parameter}, we apply the Taylor expansion for the function about $\rho_R$ and retain only the constant and first-order terms of $\rho_R$. (2) Considering the integral region $r\geq 4\eta M$, we expand the function about $r$ to spatial infinity and ignore the higher order terms than $(1/r)^{N_{\text{max}}}$.

It is also important to note that Eq.~\eqref{profile}, Eq.~\eqref{gsol} and Eq.~\eqref{fsol} only describe the scenarios in the spike region with $r\in (4\eta M, \infty)$. When $r\in (2M,4\eta M)$ and $r\rightarrow \infty$ or we have $\rho_R=0$, the density vanishes with $\rho(r)=0$ and the spacetime returns to Schwarzschild BHs with $f(r)=g(r)=1-\frac{2M}{r}$. Table~\ref{Tab:region} contains the different scenarios in both regions.

We then discuss the rationality of neglecting the pressure $p(r)$ in the spike region. Following the work of~\cite{Daghigh_2022}, Eq.~\eqref{TOV3} needs to be considered. By substituting Eq.~\eqref{profile} and Eq.~\eqref{gsol}, the pressure $p(r)$ can be numerically solved as shown in Fig.~\ref{fig:pressure}. In this way, we find it valid for $p(r) \ll \rho(r)$ and $8 \pi r^2 p(r) \ll 1-g(r)$ in the spike region.

\section{axial perturbations of Schwarzschild-like black holes}\label{sec:QNMeqn}

In this section, we consider the linear gravitational perturbation for the modified Schwarzschild-like background metric given by Eq.~\eqref{metric}. We shall deal with the axial perturbation in Sec.~\ref{sec:Axial} and the master equations in different regions are contained in Table \ref{Tab:region}. The discontinuity from the density profile results in the discontinuity of the QNMs equations (as shown in Fig.~\ref{fig:profile} and Fig.~\ref{fig:VAxial}), and the numerical method for such a problem will be discussed in Sec.~\ref{sec:Method}.

In general, the perturbation can be caused by the injection of gravitational waves or the fall of a particle into the BHs~\cite{1988sfbh.book.....F,SchFallingParticle}, in addition to describing the remnants of a binary BH merger event. Such perturbed metric can be described by:
\begin{equation}\label{LinearMetric}
    g_{\mu \nu}=\mathring{g}_{\mu \nu}+h_{\mu \nu}
\end{equation}
where $\mathring{g}_{\mu \nu}$ is the metric of background spacetime given by Eq.~\eqref{metric} and Table \ref{Tab:region}, while $h_{\mu \nu}$ is the linear perturbation term. And higher-order perturbation terms are neglected.

Because of the static spherical symmetry of the background spacetime $\mathcal{M}^4(t,r,\theta,\phi)$, it can be regarded as the product of a Lorentzian 2-dimension manifold $M^2(t,r)$ and a 2-dimension unit sphere surface manifold $S^2(\theta,\phi)$. By taking advantage of this, the metric perturbation $h_{\mu\nu}$ can be decomposed into multipoles known as axial (odd) parity and polar (even) parity as ~\cite{Thompson_2017,Nagar_2005}:
    \begin{align}
       & h_{\mu \nu}=\sum_{\ell=0}^{\infty}\sum_{m=-\ell}^{m=\ell}\left[\left(h_{\mu \nu}^{\ell m}\right)^{(\mathrm{axial})}+\left(h_{\mu \nu}^{\ell m}\right)^{(\mathrm{polar})}\right]
    \end{align}
where, $\ell$ and $m$ are the integers from the seperation of $\theta$ and $\phi$ respectively. By adopting the Regge-Wheeler (RW) gauge, the axial perturbations can be parameterized as~\cite{PhysRevD.1.2870,Berti_2009}:
    \begin{align}
      & \left(h_{\mu \nu}^{\ell m}\right)^{(\mathrm{axial})}=\left(\sin \theta \frac{\partial Y_{\ell 0}(\theta)}{\partial \theta}\right)\mathrm{e}^{i\omega t}\epsilon\cdot\nonumber\\
      &\left(\begin{array}{cccc}
        0 & 0 & 0 & h_0(r) \\
        0 & 0 & 0 & h_1(r) \\
        0 & 0 & 0 & 0 \\
        h_0(r) & h_1(r) & 0 & 0
        \end{array}\right)\label{axialCoe}
    \end{align}
and that of the polar perturbations is:
    \begin{align}
        &\left(h_{\mu \nu}^{\ell m}\right)^{(\mathrm{polar})}=Y_{\ell 0}(\theta)\mathrm{e}^{i\omega t}\epsilon\cdot\nonumber\\
        &\left(\begin{array}{cccc}
        H_0(r) (1-\frac{2M}{r}) & H_1(r) & 0 & 0 \\
        H_1(r) & \frac{H_2(r)}{1-\frac{2M}{r}} & 0 & 0 \\
        0 & 0 & r^2 K(r) & 0 \\
        0 & 0 & 0 & r^2 K(r) \sin ^2 \theta
        \end{array}\right)\label{polarCoe}
    \end{align}
where $\left| \epsilon \right| \ll 1$ is the real number to track the order of perturbation proposed by~\cite{PhysRevD.104.124082}. $\omega$ is the eigen frequency from the separation of $t$ known as the quasi-normal modes (QNMs) of the BHs. And $Y_{\ell m}$ stands for the spherical harmonics with $m=0$ for the spherical symmetry case.

Notice that the axial gravitational perturbation is intrinsically decoupled with scalar fields, the perturbations of DM will be neglected for simplification, and the corresponding perturbation equations for the axial case can be then obtained in Sec.\ref{sec:Axial}. However, such an assumption is not valid for the polar perturbations, which always couple with any extra matter fields~\cite{PhysRevLett.129.241103}. Thus, obtaining the QNMs equations for polar perturbations is a non-trivial problem, and we would like to just focus on the axial case in the rest paper.

\begin{table}
    \caption{Summary of the functions or equations in different regions. $\rho(r)$ is the DM profile, $f(r)$ and $g(r)$ are modified gauge functions in Eq.\eqref{metric} and $r_*$ is the tortoise coordinate.}
    \begin{ruledtabular}
    \begin{tabular}{c|c|c}
    & $r\in(2M,4\eta M)$ & $r\in(4\eta M, \infty)$\\
    \hline \hline
    $\rho(r)$ & $0$ & Eq.~\eqref{profile}\\
    $f(r)$ & $1-\frac{2M}{r}$ & Eq.~\eqref{fsol}\\
    $g(r)$ & $1-\frac{2M}{r}$ & Eq.~\eqref{gsol}\\
    $r_*$ & Eq.~\eqref{tortoSch} & Eq.~\eqref{tortoAxial}\\
    Axial equations & Eq.~\eqref{QNMsRW} & Eq.~\eqref{QNMsAxial}
    \end{tabular}
    \end{ruledtabular}
    \label{Tab:region}
\end{table}

\subsection{QNM Equations for Axial Perturbation}\label{sec:Axial}

Based on~\cite{liu2023gauge,PhysRevD.104.124082}, we substitute Eq.~\eqref{axialCoe} into the Einstein equations $G_{\mu\nu}=8 \pi T_{\mu\nu}$ and neglect the perturbation of DM. In the DM spike region, the QNM equations of the axial gravitational perturbation in the frequency domain can be obtained:
    \begin{equation}\label{QNMsAxial}
    \left[\frac{\partial^2}{\partial r_*^2}+\omega^2-V_{\text{axial}}(r)\right] \Psi(r)=0
    \end{equation}
where $r_*$ is the tortoise coordinate defined by:
    \begin{equation}\label{tortoAxial}
    d r_*=\frac{d r}{\sqrt{f(r) g(r)}}
    \end{equation}
and the effective potential is:
    \begin{align}
    &V_{\text{axial}}(r)=\frac{r f^{\prime}(r) g^{\prime}(r)+g(r)\left[f^{\prime}(r)+2 r f^{\prime \prime}(r)\right]}{2 r}\nonumber\\
    &-\frac{g(r) f^{\prime}(r)^2}{2 f(r)}+\frac{f(r)\left[r g^{\prime}(r)+4 g(r)+2\left(\ell^2+\ell-2\right)\right]}{2 r^2}
    \end{align}
where $g(r)$ and $f(r)$ are given by Eq.~\eqref{gsol} and Eq.~\eqref{fsol} respectively. When $r\leq 4\eta M$ or $\rho_R=0$, it returns to Schwarzschild case (see Table \ref{Tab:region}) and the QNM equations become:
    \begin{equation}\label{QNMsRW}
    \left\{\frac{\partial^2}{\partial r_*^2}+\omega^2-\left(1-\frac{2M}{r}\right)\left[\frac{\ell(\ell+1)}{r^2}-\frac{6M}{r^3}\right]\right\} \Psi=0
    \end{equation}
with $r_*$ given by:
    \begin{equation}\label{tortoSch}
    \mathrm{d} r_*=(1-\frac{2M}{r})^{-1}\mathrm{d} r
    \end{equation}
which is known as RW equation~\cite{RW}.

\subsection{Matrix Method for QNMs}\label{sec:Method}
\begin{figure}
    \includegraphics[width=\linewidth]{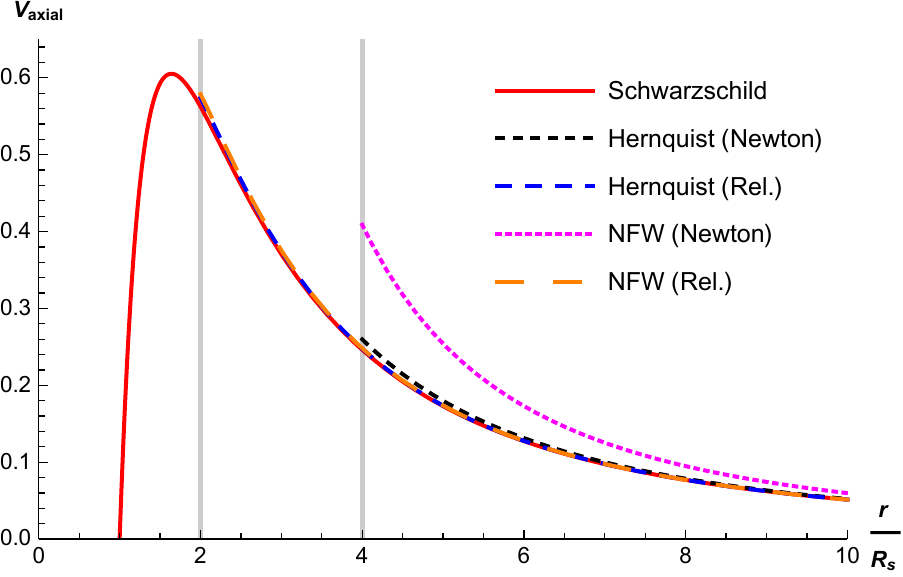}
    \caption{Comparison of the axial effective potentials for BH with and without DM. In order to make the difference obvious we use parameter $\rho_R M^2 =10^{-27}$ which is larger than  realistic values as listed in Table.~\ref{Tab:parameter}. When $r<4\eta M$, the DM vanishes and the spacetime returns to Schwarzschild, causing a discontinuous point at $r=4\eta M$.}
    \label{fig:VAxial}
\end{figure}

So far, the physical problem of perturbation has been written in the master equations as summarized in Table \ref{Tab:region}. Obviously, there is a discontinue point at $r=4\eta M$ for $\rho_R\neq 0$ as the axial effective potential is shown in Fig.~\ref{fig:VAxial} for example. And several standard methods for QNMs cannot be used directly, like WKB approximation methods~\cite{schutz1985black,PhysRevD.35.3621,PhysRevD.68.024018}. We then use the modified matrix method to deal with it~\cite{Lin_2017,lin2017matrix,https://doi.org/10.48550/arxiv.2209.11612,Shen_2022,PhysRevD.107.124002}.

We begin with the boundary conditions for QNMs of the BHs. As mentioned before, the equations return to the Schwarzschild case when $r\sim 2M$ or $r\sim \infty$. Thus, boundary conditions for QNMs are $\mathrm{e}^{-i\omega r_*}$ and $\mathrm{e}^{i\omega r_*}$, denoting the incoming waves near the horizon and outgoing waves at spatial infinity respectively~\cite{Berti_2009,https://doi.org/10.48550/arxiv.2212.00747}. The tortoise coordinate $r_*$ is defined by Eq.~\eqref{tortoSch} or can be explicitly written as $r_*=r+2M \ln(r-2M)$. And the boundary conditions now read:
    \begin{equation}\label{BC1}
        \Psi\rightarrow\begin{cases}
         \mathrm{e}^{-i\omega r}(r-2M)^{-2 i M \omega}  & r\rightarrow 2M  \\
            \mathrm{e}^{i \omega r} (r-2 M)^{2 i M \omega } & r\rightarrow\infty
            \\\end{cases}
    \end{equation}

We then introduce the substitution for $r\in (2M,4\eta M)$:
    \begin{equation}\label{changeL}
        \Psi(r)\equiv\mathrm{e}^{-i\omega r}(r-2M)^{-2 i M \omega} L(r)
    \end{equation}
    \begin{equation}\label{coorL}
        y=\frac{r-2M}{4\eta M-2M}
    \end{equation}
and for $r\in (4\eta M,\infty)$:
    \begin{equation}\label{changeR}
        \Psi(r)\equiv\mathrm{e}^{i \omega r} (r-2 M)^{2 i M \omega } R(r)
    \end{equation}
    \begin{equation}\label{coorR}
        z=1-\frac{4 \eta  M}{r}
    \end{equation}
where Eq.~\eqref{changeR} and Eq.~\eqref{changeL} remove the singularity at the boundaries, while Eq.~\eqref{coorL} and Eq.~\eqref{coorR} map $r\in (2M,4\eta M)$ and $r\in (4\eta M, \infty)$ to $[0,1]$ respectively. The QNMs equations in Table \ref{Tab:region} can be rewritten as:
    \begin{equation}
        A_2(y) L^{\prime\prime}(y) + A_1(y) L^{\prime}(y)+ A_0(y) L(y)=0
    \end{equation}
    \begin{equation}
        B_2(z) R^{\prime\prime}(z) + B_1(z) R^{\prime}(z)+ B_0(z) R(z)=0
    \end{equation}
where $A_2$, $A_1$, $A_0$, $B_2$, $B_1$ and $B_0$ are the coefficient functions about $\omega$, $y$ (or $z$) and the DM parameters in Table \ref{Tab:parameter}, depending on the axial QNMs equations. And the boundary conditions Eq.~\eqref{BC1} now become:
\begin{equation}
    L(y=0)=const,\quad R(z=1)=const
\end{equation}
By further introducing:
\begin{equation}
    \tilde{L}(y)\equiv y L(y)
\end{equation}
\begin{equation}
    \tilde{R}(z)\equiv (1-z) R(z)
\end{equation}
the boundary conditions can be further simplified as:
\begin{equation}\label{BC2}
   \tilde{L}(z=0)=\tilde{R}(z=1)=0
\end{equation}
The corresponding equations now become:
    \begin{equation}\label{eqnL}
        \tilde{A}_2(y) \tilde{L}^{\prime\prime}(y) + \tilde{A}_1(y) \tilde{L}^{\prime}(y)+ \tilde{A}_0(y) \tilde{L}(y)=0
    \end{equation}
    \begin{equation}\label{eqnR}
        \tilde{B}_2(z) \tilde{R}^{\prime\prime}(z) + \tilde{B}_1(z) \tilde{R}^{\prime}(z)+ \tilde{B}_0(z) \tilde{R}(z)=0
    \end{equation}
with
\begin{equation}
\begin{aligned}
        \tilde{A}_0(y)&=y^2 A_0\left(y\right)-y A_1\left(y\right)+2 A_2\left(y\right)\\
        \tilde{A}_1(y)&=y \left[y A_1\left(y\right)-2 A_2\left(y\right)\right]\\
        \tilde{A}_2(y)&=y^2 A_2\left(y\right)\\
        \tilde{B}_0(z)&=\left(z-1\right){}^2 B_0\left(z\right)-\left(z-1\right) B_1\left(z\right)+2 B_2\left(z\right)\\
        \tilde{B}_1(z)&=\left(z-1\right) \left[\left(z-1\right) B_1\left(z\right)-2 B_2\left(z\right)\right]\\
        \tilde{B}_2(z)&=\left(z-1\right){}^2 B_2\left(z\right)
\end{aligned}
\end{equation}

Then, we deal with the discontinuity at $r=r_c\equiv4\eta M$. Such discontinuity provides the limitation governed by Israel's junction condition~\cite{israel1966nuovo} and the wave functions on both sides are related to vanishing Wronskian~\cite{PhysRevD.59.044034,https://doi.org/10.48550/arxiv.2209.11612,Shen_2022}:
\begin{equation}
    \Psi'(r=r_c^-)\Psi(r=r_c^+)-\Psi(r=r_c^-)\Psi'(r=r_c^+)=0
\end{equation}
where $r=r_c^-$ and $r=r_c^+$ denote the asymptotic values from left or right to the discontinue point respectively. We then define the ratio coefficient $\kappa$ as:
\begin{equation}
        \kappa=\frac{\Psi'(r=r_c^-)}{\Psi(r=r_c^-)}=\frac{\Psi'(r=r_c^+)}{\Psi(r=r_c^+)}
\end{equation}
By substituting Eq.~\eqref{changeL}-Eq.~\eqref{coorR}, it can be rewritten as:
\begin{equation}\label{CCL}
\begin{aligned}
       &0=y\tilde{L}'(y)\\
       &+y \{2 (1-2 \eta ) \kappa  M y-2 i \omega  [(2 \eta -1) M y+M]-1\} \tilde{L}(y)
\end{aligned}
\end{equation}
\begin{equation}\label{CCR}
\begin{aligned}
       &0=(z-1)^2 (2 \eta +z-1) \tilde{R}'(z)\\
       &+ \left[-(2 \eta +z-1) (4 \eta  \kappa  M+z-1)+8 i \eta ^2 M \omega \right]\tilde{R}(z)
\end{aligned}
\end{equation}
And Eq.~\eqref{CCL} and Eq.~\eqref{CCR} are the connection conditions.

Then, the matrix method algorithm for our case is as follows:
\begin{itemize}
    \item (1) Following~\cite{Lin_2017}, any coordinate $x\in [0,1]$ can be discretized into $N$ points with $x_1$ to $x_{N}$. With the application of Taylor expansion, the value of a function with its 1st order derivatives to its $N$th order derivatives at each point can be written in the $N\times N$ matrices respectively (details in~\cite{Lin_2017}).
    \item (2) By substituting the matrices of the function, 1st order derivatives and 2nd order derivatives from (1), Eq.~\eqref{eqnL} and Eq.~\eqref{eqnR} can be rewritten as two matrix equations $\overline{\mathcal{M}}_L\mathcal{L}=\overline{\mathcal{M}}_R\mathcal{R}=0$ respectively. Here, $\overline{\mathcal{M}}_L$ and $\overline{\mathcal{M}}_R$ are the $N_L\times N_L$ and $N_R\times N_R$ matrices only about $\omega$ respectively, while $\mathcal{L}=\left(L(x_1), \cdots, L(x_{N_L})\right)^T$ and $\mathcal{R}=\left(R(x_1), \cdots, R(x_{N_R})\right)^T$ are the values of functions at each points.
    \item (3) Then, we use Eq.~\eqref{BC2} to replace the 1st line of $\overline{\mathcal{M}}_L$ and $N$th line of $\overline{\mathcal{M}}_R$, while we use Eq.\eqref{CCL} and Eq.~\eqref{CCR} to replace the $N$th line of $\overline{\mathcal{M}}_L$ and 1st line of $\overline{\mathcal{M}}_R$ respectively. We now obtain $\mathcal{M}_L\mathcal{L}=\mathcal{M}_R\mathcal{R}=0$, with $\mathcal{M}_L$ and $\mathcal{M}_R$ the modified matrices about $\omega$ and $\kappa$.
    \item (4) By using the equations $\det (\mathcal{M}_L)=\det (\mathcal{M}_R)=0$, the QNMs $\omega$ can be numerically solved with the corresponding ratio coefficients $\kappa$.
\end{itemize}

\section{Results}\label{sec:result}

In this section, we shall discuss the numerical results of QNMs and their characteristics affected by the DM spike. With Eq.~\eqref{rhoR}, we have unified the parameters of the DM models $r_0$ and $\rho_0$ and the parameter of the BHs $M$ into a single dimensionless parameter $\rho_R M^2$. By analyzing the parameter $\rho_R$, we can cover the analysis of the parameters $r_0$, $\rho_0$ and $M$.
We then consider the QNMs of the Sgr $\text{A}^*$ BH at the center of Milky Way in Sec.~\ref{sec:SgrA} and explore the QNMs vary with $\rho_R$ to simulate the exploration of various BHs and galaxies in the universe in Sec.~\ref{sec:QNMrhoR}. And the detectability of the deviation effected by DM spike will be discussed in Sec.~\ref{sec:detection}.

Due to the limited computational capacity, it is impossible to use arbitrarily large matrices and unlimited precision in our numerical algorithm. Based on the studies of the Mashhoon method and WKB approximate method for solving QNMs, the peak of the effective potential plays a significant role in QNMs~\cite{PhysRevLett.52.1361,schutz1985black}. And since the peak of the effective potential falls exactly in $r\leq 4\eta M$ as shown in Fig.~\ref{fig:VAxial}, in order to speed up the calculation, we choose $N_L=24$, $N_R=12$ and $N_{\text{max}}=20$.

In order to understand the computational errors that result from this choice, we first calculate the differences between the results of our algorithm with $\rho_R=0$ and the standard Schwarzschild values as listed in Table \ref{Tab:error}. And the QNMs of the Schwarzschild case from $61$ order matrix method are used as standard values, which have been proven to be accurate within 12 significant digits~\cite{https://doi.org/10.48550/arxiv.2209.11612}. Notice that the choice of $\rho_R=0$ returns the equations in Table \ref{Tab:region} back to the Schwarzschild case with corresponding QNMs, the differences in Table \ref{Tab:error} imply that our results can be credible within 6 significant digits.

\begin{table}
    \caption{The differences between the results of our algorithm with $\rho_R=0$ and the standard Schwarzschild values validate our numerical algorithm. And we use $\omega_s$ and $\omega_a$ to denote the standard Schwarzschild values and the results of the axial QNMs from our algorithm respectively.}
    \centering
    \begin{ruledtabular}
    \begin{tabular}{c | c |c }\label{Tab:error}
    $\ell$ & $n$ & $2M(\omega_{a}-\omega_{s})$  \\
    \hline \hline
     $2$ & $0$ & $-3.9\times 10^{-9} + 1.2\times10^{-8} i$ \\
     & $1$ & $-1.6\times10^{-7} + 5.6\times10^{-7} i$ \\
     $3$ & $0$ & $-8.3\times10^{-11} + 1.0\times10^{-9} i$  \\
     & $1$ & $-4.0\times10^{-8} + 7.2\times10^{-8} i$ 
    \end{tabular}
    \end{ruledtabular}
\end{table}

\subsection{The QNMs of the BH in Milky Way}\label{sec:SgrA}

We certainly want to locate a suitable observational source and target it with our GW detectors, in order to detect the ringing of a BH when it is perturbed. Due to the unsatisfactory detection of ringdown signals with ground-based GW detectors~\cite{PhysRevLett.129.111102}, we look to space-based GW detectors in the future.

Based on~\cite{PhysRevD.73.064030}, the corresponding frequency of the ringdown signal from the SMBHs with $M\sim 10^6 M_\odot$ is $\sim 10^{-2}$Hz, which falls exactly in the sensitive frequency bands of LISA. And the Sgr $\text{A}^*$ BH at the center of the Milky Way galaxy with $M=4.3\times 10^6M_\odot$ can be such a good source. The best-fit values for the parameters of the NFW model are $\rho_0=0.51 \text{GeV}/\text{cm}^3$ and $r_0=8.1 \text{kpc}$, corresponding to $\rho_R M^2=1.27\times 10^{-32}$ for NFW (Newton) and $\rho_R M^2 =5.09\times 10^{-32}$ for relativistic DM profile respectively~\cite{10.1093/mnras/stz1698}. We then calculate the QNMs of the Sgr $\text{A}^*$ BH in DM spike and list them in Table \ref{Tab:ASgrAQNM}.

As shown in Table \ref{Tab:ASgrAQNM}, the effects of the DM spike on the QNMs of the Sgr $\text{A}^*$ BH are ignorably small. The deviations of the fundamental mode from the one without DM are about $\sim10^{-9}$Hz and $\sim0.001s$. The detection of such deviations is not possible in the near future.

\begin{table*}
    \caption{We list the axial QNMs $2M \omega$ of the Sgr $\text{A}^*$ BH at the center of Milky Way galaxy surrounded by the NFW DM spike.}
    \begin{ruledtabular}
    \begin{tabular}{c c |c | c | c}\label{Tab:ASgrAQNM}
    $\ell$ & $n$ & Newton & Rel. & Schwarzschild \\
    \hline \hline
     $2$ & $0$ & $0.747342 - 0.177926 i$ & $0.747343 - 0.177924 i$ & $0.747343 - 0.177925 i$\\
     & $1$ & $0.693541 - 0.547762 i$ & $0.693422 - 0.547827 i$ & $0.693422 - 0.547829 i$\\
     $3$ & $0$ & $1.198890 - 0.185406 i$ & $1.198886 - 0.185406 i$ & $1.198887 - 0.185406 i$ \\
     & $1$ & $1.165002 - 0.562909 i$ & $1.165286 - 0.562594 i$ & $1.165288 - 0.562596 i$
    \end{tabular}
    \end{ruledtabular}
\end{table*}

\subsection{The Fundamental QNM with Varying $\rho_R$}\label{sec:QNMrhoR}

We then explore a wider range of observational sources with varying parameters. For given parameters $r_0$, $\rho_0$ and $M$, the parameter $\rho_R$ is uniquely determined by Eq.~\eqref{rhoR}. Through the exploration of the trend of QNMs with varying $\rho_R$, the corresponding trend for parameters $r_0$, $\rho_0$ and $M$ can also be obtained. Specifically, the dimensionless parameter $\rho_R M^2$ is closely related to $r_0$, $\rho$ and $M$. This exploration is interesting for three reasons~\cite{PhysRevD.104.124082,ZHANG2022101078}:

\begin{itemize}
    \item (1) The DM parameters $r_0$ and $\rho_0$ are generally derived by fitting the corresponding density profiles with the data of rotation curves in various galaxies~\cite{10.1093/mnras/stz1698}. They can accurately reflect the distribution of DM throughout the entire galaxy, but are essentially effected by BHs and basically free near BHs.
    \item (2) Around a BH, the baryonic component also makes significant contributions to the DM parameters $r_0$ and $\rho_0$.
    \item (3) The parameters $r_0$ and $\rho_0$ change from galaxy to galaxy, while $M$ changes along with BHs. A suitable BH in the DM spike is easier for detection.
\end{itemize}

\begin{figure*}
    \subfloat{
            \includegraphics[width=0.45\linewidth]{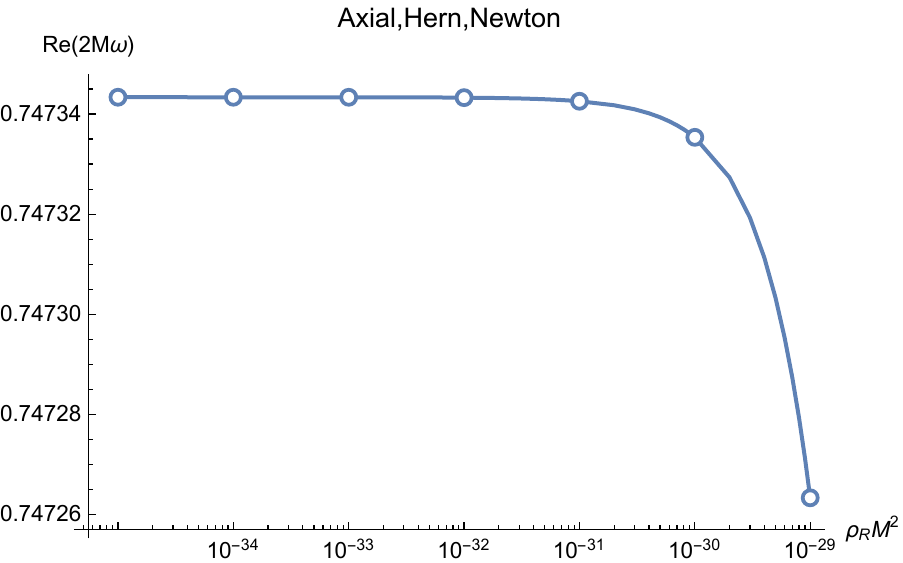}
       }
        \hfill
    \subfloat{
            \includegraphics[width=0.45\linewidth]{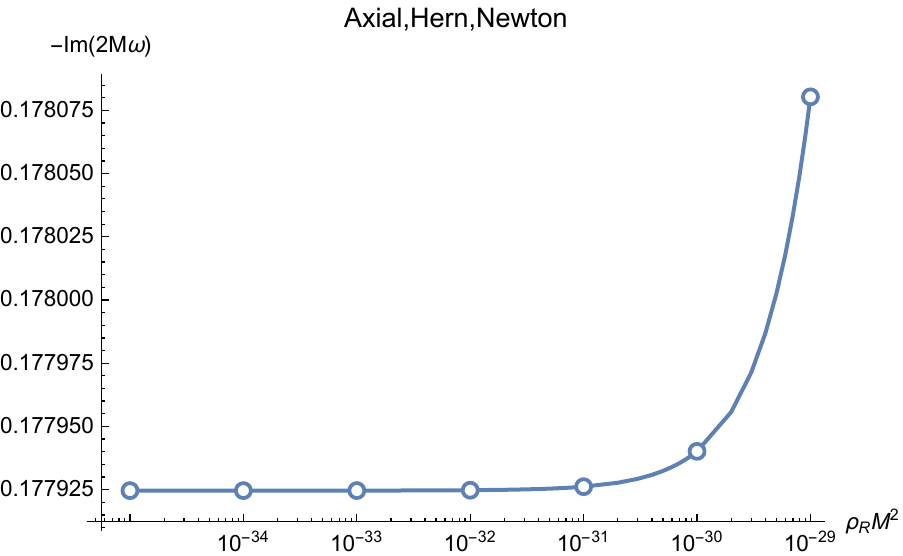}
       }
        \hfill
    \subfloat{
            \includegraphics[width=0.45\linewidth]{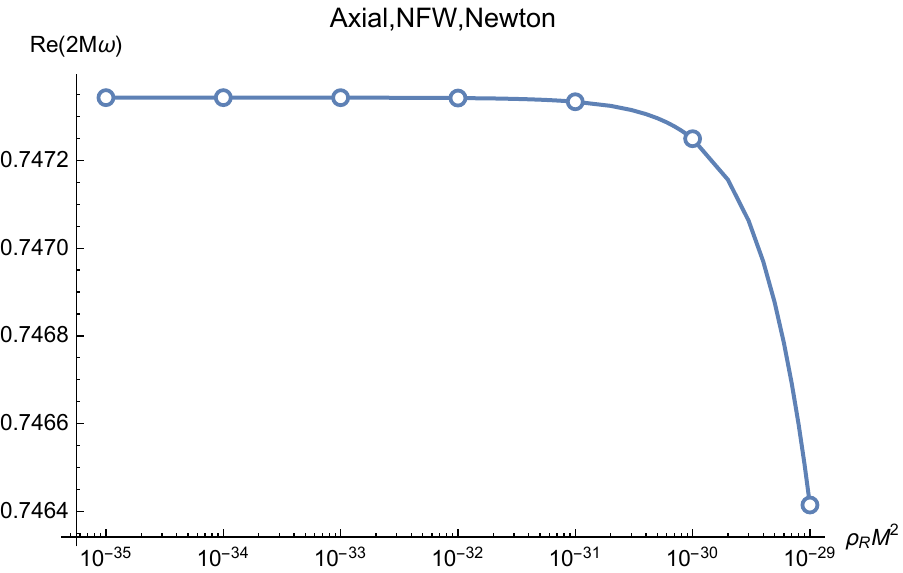}
       }
        \hfill
    \subfloat{
            \includegraphics[width=0.45\linewidth]{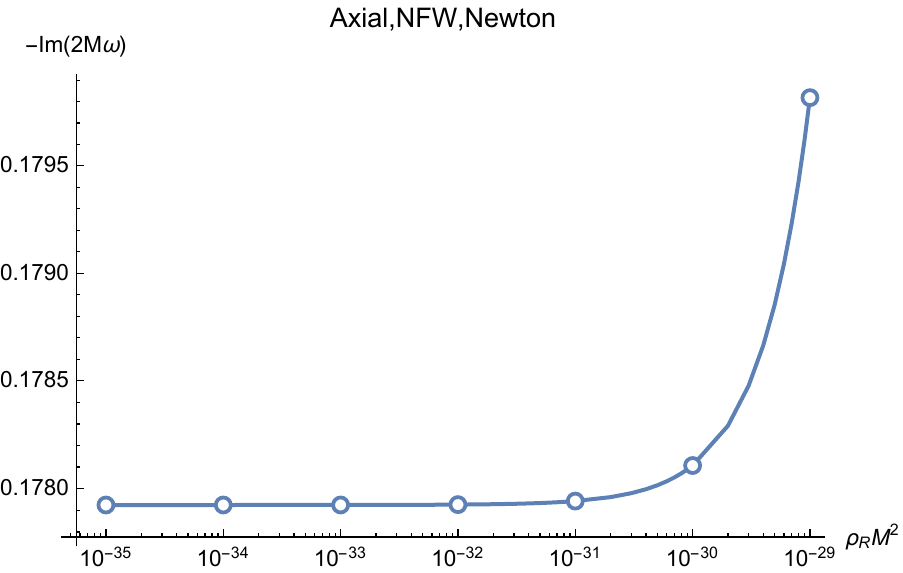}
       }
        \hfill
    \subfloat{
            \includegraphics[width=0.45\linewidth]{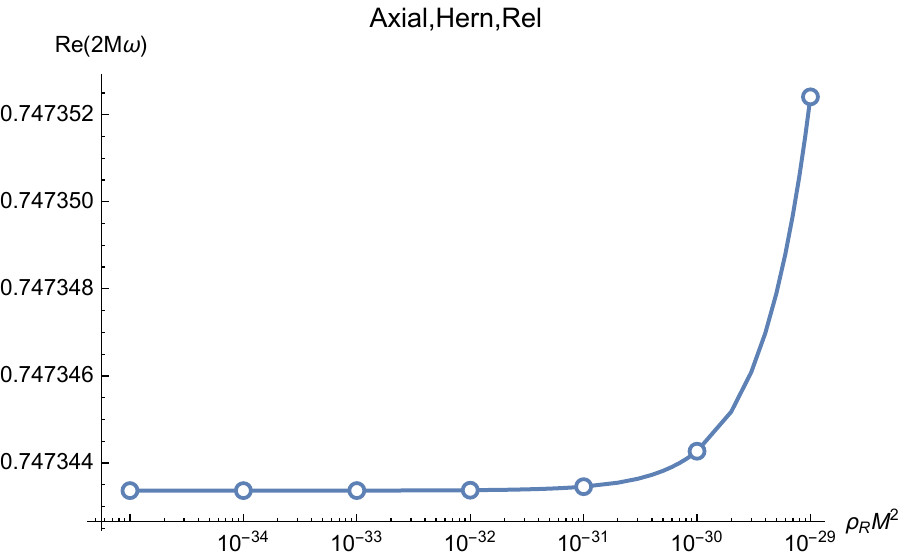}
       }
        \hfill
    \subfloat{
            \includegraphics[width=0.45\linewidth]{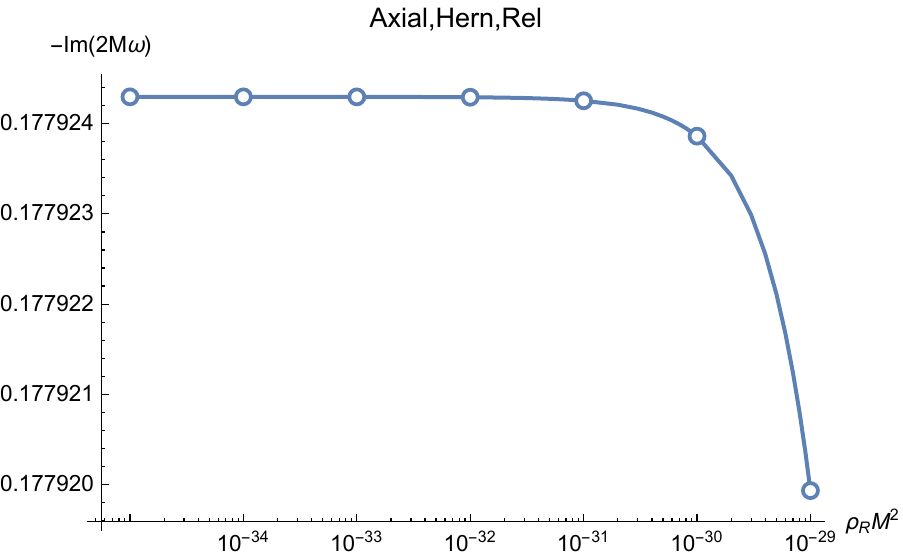}
       }
 
    \subfloat{
            \includegraphics[width=0.45\linewidth]{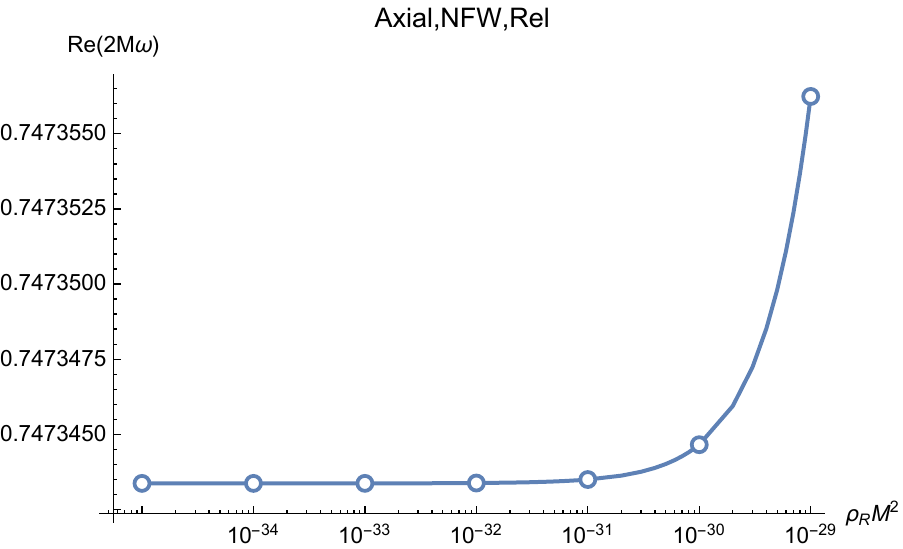}
       }
        \hfill
    \subfloat{
            \includegraphics[width=0.45\linewidth]{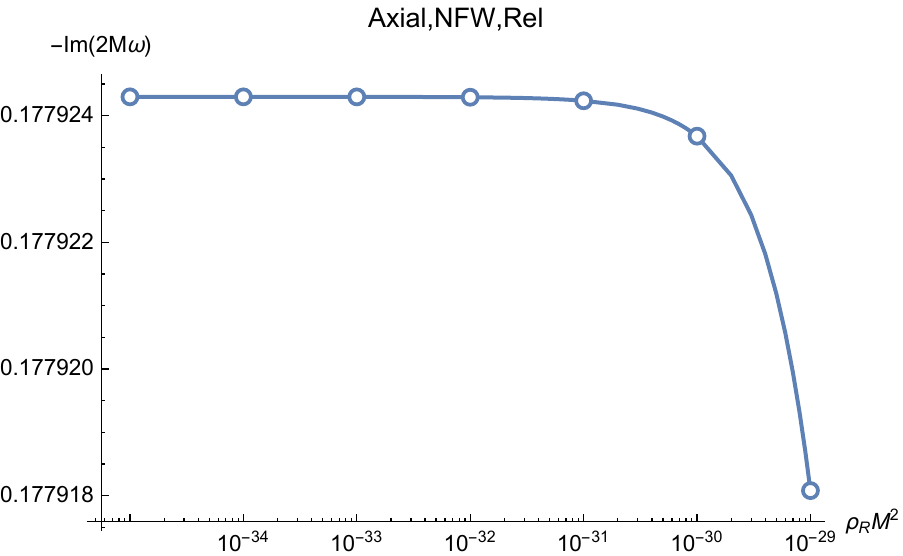}
       }

    \caption{The behavior of the QNMs with respect to the dimensionless parameter $\rho_R M^2$ for Hernquist models of DM profile. In figures, ``$\operatorname{Re}$" in the left column denotes the real part,  and ``$\operatorname{Im}$" in the right column denotes the imaginary part. The marked points correspond to $\rho_R M^2=10^{-35},10^{-34},\cdots,10^{-29}$.}
    \label{fig:QNMs_rhoR}
\end{figure*}

As shown in Fig.~\ref{fig:QNMs_rhoR}, the axial QNMs for both Hernquist and NFW models show different trends between Newton and the relativistic case. Without relativistic modification, both the real and imaginary parts of the QNMs decrease with the increasing $\rho_R M^2$. However, with relativistic modification, both the real and imaginary parts of the QNMs increase as $\rho_R M^2$ grows.


\subsection{Hints of Detectability by Space-based Detectors}\label{sec:detection}

With the previous exploration, we have found that the DM spike can result in deviations of QNMs from that of BH without DM. We then explore the possibility of detection of such deviations by future space-based detectors. The GW waveform during the ringdown process can be written as~\cite{PhysRevD.73.064030}:
\begin{equation}
    h_{+}+i h_{\times}=\frac{M_z}{D_{\mathrm{L}}} \sum_{\ell m n} \mathcal{A}_{\ell m n} e^{i\left(f_{\ell m n} t+\phi_{\ell m n}\right)} e^{-t / \tau_{\ell m n}} S_{\ell m n}
\end{equation}
where $M_z$ is the red-shifted BH mass, $D_L$ is the luminosity distance to the source, $\mathcal{A}_{\ell m n}$ is the amplitude of the corresponding QNM, $\phi_{\ell m n}$ is the phase coefficient and $S_{\ell}$ is the 2-spin-weighted spheroidal harmonics depending on the polar and azimuthal angles. And the real waveform is the superposition of the axial and polar parity components. The two parameters associated to QNMs are the GW frequency $f_{\ell m n}$ and the damping time $\tau_{\ell m n}$ defined as:
\begin{align}
&2 \pi f_{\ell m n}=\operatorname{Re}\left(\omega_{\ell m n}\right),\\
&\tau_{\ell m n}=-\frac{1}{\operatorname{Im}\left(\omega_{\ell m n}\right)},
\end{align}
where $\omega_{\ell m n}$ is the QNMs for given $(\ell,m,n)$. Here we consider only the fundamental mode with $(\ell,m,n)=(2,0,0)$ because it decays slowest. Following~\cite{ZHANG2022101078}, the frequency and the damping time can be expanded as:
\begin{align}
&f_{\ell m n}=f_{\ell m n}^{\operatorname{Sch}}\left(1+\delta f_{\ell m n}\right),\\
&\tau_{\ell m n}=\tau_{\ell m n}^{\mathrm{Sch}}\left(1+\delta \tau_{\ell m n}\right),
\end{align}
where $f_{\ell m n}^{\operatorname{Sch}}$ and $\tau_{\ell m n}^{\mathrm{Sch}}$ are the QNM frequency and damping time for Schwarzschild case, while $\delta f_{\ell m n}$ and $\delta\tau_{\ell m n}$ are the corresponding relative deviations. Based on \cite{PhysRevD.100.044036}, the relative deviations larger than $\sim 10^{-3}$ have the possibility to be detected by future space-based detectors.

\begin{figure*}
    \subfloat{
            \includegraphics[width=0.45\linewidth]{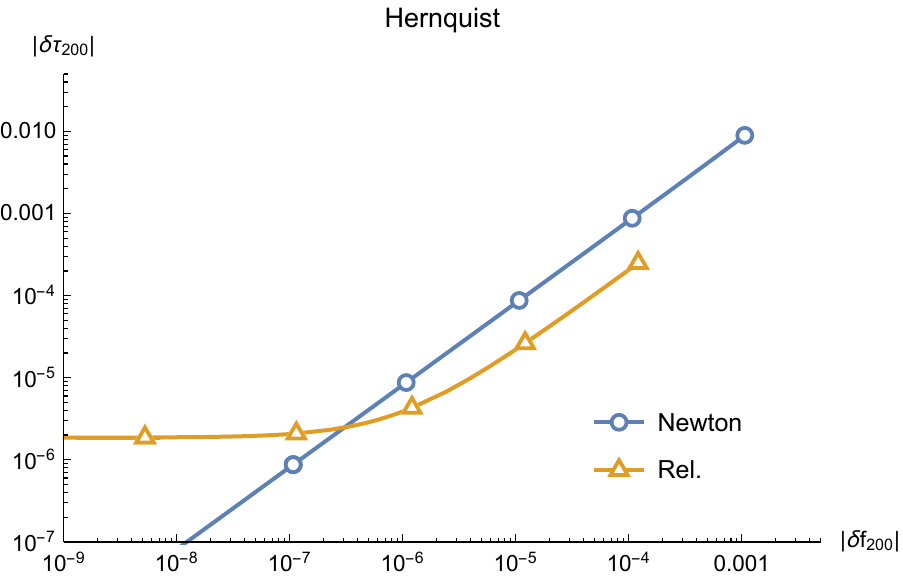}
       }
        \hfill
    \subfloat{
            \includegraphics[width=0.45\linewidth]{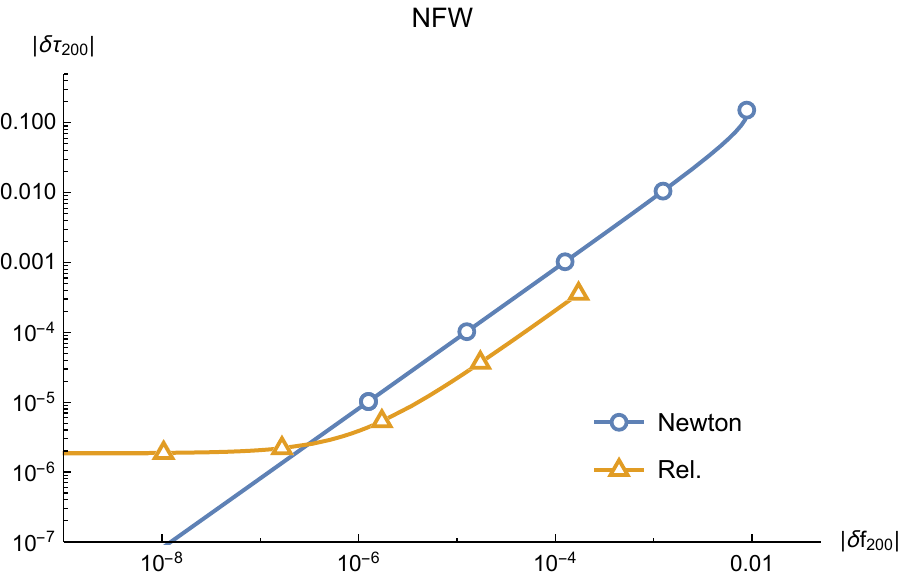}
       }

    \caption{The behavior of the deviations with respect to the dimensionless parameter $\rho_R M^2$ for different models. The points in each figure from the left to the right correspond to $\rho_R M^2=10^{-32},10^{-31},\cdots,10^{-28}$ respectively.}
    \label{fig:detection}
\end{figure*}

From Fig.\ref{fig:detection}, we can see that the deviations of axial QNMs increase along with the increasing $\rho_R M^2$. In general, the deviation for the NFW DM profile is larger than that of the Hernquist profile. The deviation for the Newtonian profile is larger than that of the relativistic profile. Corresponding to the DM profiles listed in Table.~\ref{Tab:parameter} the largest deviation of $\delta f_{\ell m n}$ corresponding to $\rho_R M^2\sim10^{-28}$ and with NFW DM profile may be as high as $\sim 10^{-2}$ without relativistic modification and about $\sim 10^{-4}$ with relativistic modification. In the meantime, $\delta \tau_{\ell m n}$ can reach $0.1$ and $0.001$ respectively. Such deviations have the possibility to be detected by future space-based detectors~\cite{PhysRevD.100.044036}.

\section{Conclusion}

In this paper, we explore the behavior of the GWs waveform emitted by the perturbed Schwarzschild-like BHs in the DM spike. Accounting for the profile of the DM spike we use the TOV equation to derive the modified metrics. After that, we derive the QNMs equations for axial case under the modified metrics and use modified matrix methods to calculate the QNMs.
Finally, we discuss the impacts of the DM spike and analyze the detectability.

The key points of the exploration are as follows:
\begin{itemize}
    \item (1) We unified the parameters $r_0$, $\rho_0$ and $M$ into a single dimensionless parameter $\rho_R M^2$ by Eq.~\eqref{rhoR}. And the analysis of $\rho_R M^2$ can cover all three parameters.
    \item (2) We neglect the pressure $p(r)$ in the DM spike and show the  rationality in Fig.~\ref{fig:pressure}.
    \item (3) The DM spike profile vanishes for $r\leq 4\eta M$ which divides the entire domain into two different regions as shown in Fig.~\ref{fig:profile}. And the behaviors of the spacetime and QNMs equations differ in both regions as shown in Table \ref{Tab:region}.
    \item (4) Such differences in both regions lead to the discontinuity of QNMs equations at $4 \eta M$ as shown in Fig.~\ref{fig:VAxial}. And the modified matrix method for such a scenario is used.
    \item (5) We choose $N_L=24$, $N_R=12$ and $N_{\text{max}}=20$, and the numerical results are as accurate as 6 significant digits as shown in Table \ref{Tab:error}.
    \item (6) The QNMs of the Sgr $\text{A}^*$ BH at the center of the Milky Way galaxy in the NFW DM spike are calculated and listed in Table \ref{Tab:ASgrAQNM}.
    \item (7) We explore the trend of fundamental QNM with varying $\rho_R$ to simulate the varying BHs and galaxies. The trends with and without relativistic modification are different: both the real and imaginary parts decrease as $\rho_R M^2$ increase without relativistic modification, while they increase with relativistic modification. And details are shown in Fig.~\ref{fig:QNMs_rhoR}.
     \item (8) The detectability of the deviations affected by the DM spike is explored. The relativistic modification cause less relative deviation than Newton case. For the DM profiles listed in Table.~\ref{Tab:parameter}, the relative deviation of the QNM frequency can be as high as $\sim10^{-2}$, which can be detected by future space-based detectors.
\end{itemize}

In summary, the presence of the DM spike does impact the axial ringing GW waveform of BHs. Only spinless BH with non-self-interaction DM spike is investigated in the current paper. The self-interaction DM may change the DM profile, and the effect of the DM spike for spinning BH may be even stronger because the spin of a BH can enhance the spike effect~\cite{PhysRevD.96.083014} and enlarge the decay time of QNMs. And these will be the focus of our next exploration.

Another point of interest, of course, is the impact on QNMs of polar case. As is mentioned in Sec.~\ref{sec:QNMeqn}, the perturbations of DM always couple with any extra matter fields~\cite{PhysRevLett.129.241103}, which may have a more significant impact on polar QNMs and then easier to be detected. And this will also be the focus of our next exploration.


\acknowledgements


The authors would like to thank Dr. Zhan-Feng Mai and Prof. Run-Qiu Yang for discussing and providing useful ideas to calculate QNMs in this work.

This work was supported in part by the National Key Research and Development Program of China Grant No. 2021YFC2203001 and in part by the NSFC (No.~11920101003, No.~12021003 and No.~12005016). Z. Cao was supported by ``the Interdiscipline Research Funds of Beijing Normal University" and CAS Project for Young Scientists in Basic Research YSBR-006.

\bibliography{reference}

\begin{thebibliography}{71}%
\makeatletter
\providecommand \@ifxundefined [1]{%
 \@ifx{#1\undefined}
}%
\providecommand \@ifnum [1]{%
 \ifnum #1\expandafter \@firstoftwo
 \else \expandafter \@secondoftwo
 \fi
}%
\providecommand \@ifx [1]{%
 \ifx #1\expandafter \@firstoftwo
 \else \expandafter \@secondoftwo
 \fi
}%
\providecommand \natexlab [1]{#1}%
\providecommand \enquote  [1]{``#1''}%
\providecommand \bibnamefont  [1]{#1}%
\providecommand \bibfnamefont [1]{#1}%
\providecommand \citenamefont [1]{#1}%
\providecommand \href@noop [0]{\@secondoftwo}%
\providecommand \href [0]{\begingroup \@sanitize@url \@href}%
\providecommand \@href[1]{\@@startlink{#1}\@@href}%
\providecommand \@@href[1]{\endgroup#1\@@endlink}%
\providecommand \@sanitize@url [0]{\catcode `\\12\catcode `\$12\catcode
  `\&12\catcode `\#12\catcode `\^12\catcode `\_12\catcode `\%12\relax}%
\providecommand \@@startlink[1]{}%
\providecommand \@@endlink[0]{}%
\providecommand \url  [0]{\begingroup\@sanitize@url \@url }%
\providecommand \@url [1]{\endgroup\@href {#1}{\urlprefix }}%
\providecommand \urlprefix  [0]{URL }%
\providecommand \Eprint [0]{\href }%
\providecommand \doibase [0]{https://doi.org/}%
\providecommand \selectlanguage [0]{\@gobble}%
\providecommand \bibinfo  [0]{\@secondoftwo}%
\providecommand \bibfield  [0]{\@secondoftwo}%
\providecommand \translation [1]{[#1]}%
\providecommand \BibitemOpen [0]{}%
\providecommand \bibitemStop [0]{}%
\providecommand \bibitemNoStop [0]{.\EOS\space}%
\providecommand \EOS [0]{\spacefactor3000\relax}%
\providecommand \BibitemShut  [1]{\csname bibitem#1\endcsname}%
\let\auto@bib@innerbib\@empty
\bibitem [{\citenamefont {{Einstein}}(1915)}]{einstein1915feldgleichungen}%
  \BibitemOpen
  \bibfield  {author} {\bibinfo {author} {\bibfnamefont {A.}~\bibnamefont
  {{Einstein}}},\ }\bibfield  {title} {\bibinfo {title} {{Die Feldgleichungen
  der Gravitation}},\ }\href@noop {} {\bibfield  {journal} {\bibinfo  {journal}
  {Sitzungsberichte der K\&ouml;niglich Preussischen Akademie der
  Wissenschaften}\ ,\ \bibinfo {pages} {844}} (\bibinfo {year}
  {1915})}\BibitemShut {NoStop}%
\bibitem [{\citenamefont {Askar}\ \emph {et~al.}(2019)\citenamefont {Askar},
  \citenamefont {Belczynski}, \citenamefont {Bertone}, \citenamefont {Bon},
  \citenamefont {Blas}, \citenamefont {Brito}, \citenamefont {Bulik},
  \citenamefont {Burrage}, \citenamefont {Byrnes}, \citenamefont {Caprini}
  \emph {et~al.}}]{Barack_2019}%
  \BibitemOpen
  \bibfield  {author} {\bibinfo {author} {\bibfnamefont {A.}~\bibnamefont
  {Askar}}, \bibinfo {author} {\bibfnamefont {C.}~\bibnamefont {Belczynski}},
  \bibinfo {author} {\bibfnamefont {G.}~\bibnamefont {Bertone}}, \bibinfo
  {author} {\bibfnamefont {E.}~\bibnamefont {Bon}}, \bibinfo {author}
  {\bibfnamefont {D.}~\bibnamefont {Blas}}, \bibinfo {author} {\bibfnamefont
  {R.}~\bibnamefont {Brito}}, \bibinfo {author} {\bibfnamefont
  {T.}~\bibnamefont {Bulik}}, \bibinfo {author} {\bibfnamefont
  {C.}~\bibnamefont {Burrage}}, \bibinfo {author} {\bibfnamefont {C.~T.}\
  \bibnamefont {Byrnes}}, \bibinfo {author} {\bibfnamefont {C.}~\bibnamefont
  {Caprini}}, \emph {et~al.},\ }\bibfield  {title} {\bibinfo {title} {Black
  holes, gravitational waves and fundamental physics: a roadmap},\ }\href
  {https://doi.org/10.1088/1361-6382/ab0587} {\bibfield  {journal} {\bibinfo
  {journal} {Classical and Quantum Gravity}\ }\textbf {\bibinfo {volume}
  {36}},\ \bibinfo {pages} {143001} (\bibinfo {year} {2019})}\BibitemShut
  {NoStop}%
\bibitem [{\citenamefont {Abbott}\ \emph {et~al.}(2016)\citenamefont {Abbott},
  \citenamefont {Abbott}, \citenamefont {Abbott}, \citenamefont {Abernathy},
  \citenamefont {Acernese}, \citenamefont {Ackley}, \citenamefont {Adams},
  \citenamefont {Adams}, \citenamefont {Addesso}, \citenamefont {Adhikari}
  \emph {et~al.}}]{PhysRevLett.116.061102}%
  \BibitemOpen
  \bibfield  {author} {\bibinfo {author} {\bibfnamefont {B.~P.}\ \bibnamefont
  {Abbott}}, \bibinfo {author} {\bibfnamefont {R.}~\bibnamefont {Abbott}},
  \bibinfo {author} {\bibfnamefont {T.~D.}\ \bibnamefont {Abbott}}, \bibinfo
  {author} {\bibfnamefont {M.~R.}\ \bibnamefont {Abernathy}}, \bibinfo {author}
  {\bibfnamefont {F.}~\bibnamefont {Acernese}}, \bibinfo {author}
  {\bibfnamefont {K.}~\bibnamefont {Ackley}}, \bibinfo {author} {\bibfnamefont
  {C.}~\bibnamefont {Adams}}, \bibinfo {author} {\bibfnamefont
  {T.}~\bibnamefont {Adams}}, \bibinfo {author} {\bibfnamefont
  {P.}~\bibnamefont {Addesso}}, \bibinfo {author} {\bibfnamefont {R.~X.}\
  \bibnamefont {Adhikari}}, \emph {et~al.} (\bibinfo {collaboration} {LIGO
  Scientific Collaboration and Virgo Collaboration}),\ }\bibfield  {title}
  {\bibinfo {title} {Observation of gravitational waves from a binary black
  hole merger},\ }\href {https://doi.org/10.1103/PhysRevLett.116.061102}
  {\bibfield  {journal} {\bibinfo  {journal} {Phys. Rev. Lett.}\ }\textbf
  {\bibinfo {volume} {116}},\ \bibinfo {pages} {061102} (\bibinfo {year}
  {2016})}\BibitemShut {NoStop}%
\bibitem [{GWO(2022)}]{GWOSC}%
  \BibitemOpen
  \href@noop {} {\bibinfo {title} {Gravitational wave open science center}},\
  \bibinfo {howpublished}
  {\url{https://www.gw-openscience.org/eventapi/html/allevents/}} (\bibinfo
  {year} {2022})\BibitemShut {NoStop}%
\bibitem [{\citenamefont {Abbott}\ \emph {et~al.}(2019)\citenamefont {Abbott},
  \citenamefont {Abbott}, \citenamefont {Abbott}, \citenamefont {Abraham},
  \citenamefont {Acernese}, \citenamefont {Ackley}, \citenamefont {Adams},
  \citenamefont {Adhikari}, \citenamefont {Adya}, \citenamefont {Affeldt} \emph
  {et~al.}}]{GWTC1}%
  \BibitemOpen
  \bibfield  {author} {\bibinfo {author} {\bibfnamefont {B.~P.}\ \bibnamefont
  {Abbott}}, \bibinfo {author} {\bibfnamefont {R.}~\bibnamefont {Abbott}},
  \bibinfo {author} {\bibfnamefont {T.~D.}\ \bibnamefont {Abbott}}, \bibinfo
  {author} {\bibfnamefont {S.}~\bibnamefont {Abraham}}, \bibinfo {author}
  {\bibfnamefont {F.}~\bibnamefont {Acernese}}, \bibinfo {author}
  {\bibfnamefont {K.}~\bibnamefont {Ackley}}, \bibinfo {author} {\bibfnamefont
  {C.}~\bibnamefont {Adams}}, \bibinfo {author} {\bibfnamefont {R.~X.}\
  \bibnamefont {Adhikari}}, \bibinfo {author} {\bibfnamefont {V.~B.}\
  \bibnamefont {Adya}}, \bibinfo {author} {\bibfnamefont {C.}~\bibnamefont
  {Affeldt}}, \emph {et~al.} (\bibinfo {collaboration} {LIGO Scientific
  Collaboration and Virgo Collaboration}),\ }\bibfield  {title} {\bibinfo
  {title} {Gwtc-1: A gravitational-wave transient catalog of compact binary
  mergers observed by ligo and virgo during the first and second observing
  runs},\ }\href {https://doi.org/10.1103/PhysRevX.9.031040} {\bibfield
  {journal} {\bibinfo  {journal} {Phys. Rev. X}\ }\textbf {\bibinfo {volume}
  {9}},\ \bibinfo {pages} {031040} (\bibinfo {year} {2019})}\BibitemShut
  {NoStop}%
\bibitem [{\citenamefont {Abbott}\ \emph
  {et~al.}(2021{\natexlab{a}})\citenamefont {Abbott}, \citenamefont {Abbott},
  \citenamefont {Abraham}, \citenamefont {Acernese}, \citenamefont {Ackley},
  \citenamefont {Adams}, \citenamefont {Adams}, \citenamefont {Adhikari},
  \citenamefont {Adya}, \citenamefont {Affeldt} \emph {et~al.}}]{GWTC2}%
  \BibitemOpen
  \bibfield  {author} {\bibinfo {author} {\bibfnamefont {R.}~\bibnamefont
  {Abbott}}, \bibinfo {author} {\bibfnamefont {T.~D.}\ \bibnamefont {Abbott}},
  \bibinfo {author} {\bibfnamefont {S.}~\bibnamefont {Abraham}}, \bibinfo
  {author} {\bibfnamefont {F.}~\bibnamefont {Acernese}}, \bibinfo {author}
  {\bibfnamefont {K.}~\bibnamefont {Ackley}}, \bibinfo {author} {\bibfnamefont
  {A.}~\bibnamefont {Adams}}, \bibinfo {author} {\bibfnamefont
  {C.}~\bibnamefont {Adams}}, \bibinfo {author} {\bibfnamefont {R.~X.}\
  \bibnamefont {Adhikari}}, \bibinfo {author} {\bibfnamefont {V.~B.}\
  \bibnamefont {Adya}}, \bibinfo {author} {\bibfnamefont {C.}~\bibnamefont
  {Affeldt}}, \emph {et~al.} (\bibinfo {collaboration} {LIGO Scientific
  Collaboration and Virgo Collaboration}),\ }\bibfield  {title} {\bibinfo
  {title} {Gwtc-2: Compact binary coalescences observed by ligo and virgo
  during the first half of the third observing run},\ }\href
  {https://doi.org/10.1103/PhysRevX.11.021053} {\bibfield  {journal} {\bibinfo
  {journal} {Phys. Rev. X}\ }\textbf {\bibinfo {volume} {11}},\ \bibinfo
  {pages} {021053} (\bibinfo {year} {2021}{\natexlab{a}})}\BibitemShut
  {NoStop}%
\bibitem [{\citenamefont {Abbott}\ \emph
  {et~al.}(2021{\natexlab{b}})\citenamefont {Abbott}, \citenamefont {Abbott},
  \citenamefont {Acernese}, \citenamefont {Ackley}, \citenamefont {Adams},
  \citenamefont {Adhikari}, \citenamefont {Adhikari}, \citenamefont {Adya},
  \citenamefont {Affeldt}, \citenamefont {Agarwal} \emph {et~al.}}]{GWTC2.1}%
  \BibitemOpen
  \bibfield  {author} {\bibinfo {author} {\bibfnamefont {R.}~\bibnamefont
  {Abbott}}, \bibinfo {author} {\bibfnamefont {T.}~\bibnamefont {Abbott}},
  \bibinfo {author} {\bibfnamefont {F.}~\bibnamefont {Acernese}}, \bibinfo
  {author} {\bibfnamefont {K.}~\bibnamefont {Ackley}}, \bibinfo {author}
  {\bibfnamefont {C.}~\bibnamefont {Adams}}, \bibinfo {author} {\bibfnamefont
  {N.}~\bibnamefont {Adhikari}}, \bibinfo {author} {\bibfnamefont
  {R.}~\bibnamefont {Adhikari}}, \bibinfo {author} {\bibfnamefont
  {V.}~\bibnamefont {Adya}}, \bibinfo {author} {\bibfnamefont {C.}~\bibnamefont
  {Affeldt}}, \bibinfo {author} {\bibfnamefont {D.}~\bibnamefont {Agarwal}},
  \emph {et~al.},\ }\bibfield  {title} {\bibinfo {title} {Gwtc-2.1: Deep
  extended catalog of compact binary coalescences observed by ligo and virgo
  during the first half of the third observing run},\ }\bibfield  {journal}
  {\bibinfo  {journal} {arXiv preprint arXiv:2108.01045}\ }\href
  {https://doi.org/10.48550/ARXIV.2108.01045} {10.48550/ARXIV.2108.01045}
  (\bibinfo {year} {2021}{\natexlab{b}})\BibitemShut {NoStop}%
\bibitem [{\citenamefont {Abbott}\ \emph
  {et~al.}(2021{\natexlab{c}})\citenamefont {Abbott}, \citenamefont {Abbott},
  \citenamefont {Acernese}, \citenamefont {Ackley}, \citenamefont {Adams},
  \citenamefont {Adhikari}, \citenamefont {Adhikari}, \citenamefont {Adya},
  \citenamefont {Affeldt}, \citenamefont {Agarwal} \emph {et~al.}}]{GWTC3}%
  \BibitemOpen
  \bibfield  {author} {\bibinfo {author} {\bibfnamefont {R.}~\bibnamefont
  {Abbott}}, \bibinfo {author} {\bibfnamefont {T.}~\bibnamefont {Abbott}},
  \bibinfo {author} {\bibfnamefont {F.}~\bibnamefont {Acernese}}, \bibinfo
  {author} {\bibfnamefont {K.}~\bibnamefont {Ackley}}, \bibinfo {author}
  {\bibfnamefont {C.}~\bibnamefont {Adams}}, \bibinfo {author} {\bibfnamefont
  {N.}~\bibnamefont {Adhikari}}, \bibinfo {author} {\bibfnamefont
  {R.}~\bibnamefont {Adhikari}}, \bibinfo {author} {\bibfnamefont
  {V.}~\bibnamefont {Adya}}, \bibinfo {author} {\bibfnamefont {C.}~\bibnamefont
  {Affeldt}}, \bibinfo {author} {\bibfnamefont {D.}~\bibnamefont {Agarwal}},
  \emph {et~al.},\ }\bibfield  {title} {\bibinfo {title} {Gwtc-3: compact
  binary coalescences observed by ligo and virgo during the second part of the
  third observing run},\ }\bibfield  {journal} {\bibinfo  {journal} {arXiv
  preprint arXiv:2111.03606}\ }\href
  {https://doi.org/10.48550/ARXIV.2111.03606} {10.48550/ARXIV.2111.03606}
  (\bibinfo {year} {2021}{\natexlab{c}})\BibitemShut {NoStop}%
\bibitem [{\citenamefont {Collaboration}\ \emph
  {et~al.}(2019{\natexlab{a}})\citenamefont {Collaboration}, \citenamefont
  {Akiyama}, \citenamefont {Alberdi}, \citenamefont {Alef}, \citenamefont
  {Asada}, \citenamefont {Azulay}, \citenamefont {Baczko}, \citenamefont
  {Ball}, \citenamefont {Baloković}, \citenamefont {Barrett}, \citenamefont
  {Bintley} \emph {et~al.}}]{1Akiyama_2019}%
  \BibitemOpen
  \bibfield  {author} {\bibinfo {author} {\bibfnamefont {T.~E. H.~T.}\
  \bibnamefont {Collaboration}}, \bibinfo {author} {\bibfnamefont
  {K.}~\bibnamefont {Akiyama}}, \bibinfo {author} {\bibfnamefont
  {A.}~\bibnamefont {Alberdi}}, \bibinfo {author} {\bibfnamefont
  {W.}~\bibnamefont {Alef}}, \bibinfo {author} {\bibfnamefont {K.}~\bibnamefont
  {Asada}}, \bibinfo {author} {\bibfnamefont {R.}~\bibnamefont {Azulay}},
  \bibinfo {author} {\bibfnamefont {A.-K.}\ \bibnamefont {Baczko}}, \bibinfo
  {author} {\bibfnamefont {D.}~\bibnamefont {Ball}}, \bibinfo {author}
  {\bibfnamefont {M.}~\bibnamefont {Baloković}}, \bibinfo {author}
  {\bibfnamefont {J.}~\bibnamefont {Barrett}}, \bibinfo {author} {\bibfnamefont
  {D.}~\bibnamefont {Bintley}}, \emph {et~al.},\ }\bibfield  {title} {\bibinfo
  {title} {First m87 event horizon telescope results. i. the shadow of the
  supermassive black hole},\ }\href {https://doi.org/10.3847/2041-8213/ab0ec7}
  {\bibfield  {journal} {\bibinfo  {journal} {The Astrophysical Journal
  Letters}\ }\textbf {\bibinfo {volume} {875}},\ \bibinfo {pages} {L1}
  (\bibinfo {year} {2019}{\natexlab{a}})}\BibitemShut {NoStop}%
\bibitem [{\citenamefont {Collaboration}\ \emph
  {et~al.}(2019{\natexlab{b}})\citenamefont {Collaboration}, \citenamefont
  {Akiyama}, \citenamefont {Alberdi}, \citenamefont {Alef}, \citenamefont
  {Asada}, \citenamefont {Azulay}, \citenamefont {Baczko}, \citenamefont
  {Ball}, \citenamefont {Baloković}, \citenamefont {Barrett}, \citenamefont
  {Bintley} \emph {et~al.}}]{2Akiyama_2019}%
  \BibitemOpen
  \bibfield  {author} {\bibinfo {author} {\bibfnamefont {T.~E. H.~T.}\
  \bibnamefont {Collaboration}}, \bibinfo {author} {\bibfnamefont
  {K.}~\bibnamefont {Akiyama}}, \bibinfo {author} {\bibfnamefont
  {A.}~\bibnamefont {Alberdi}}, \bibinfo {author} {\bibfnamefont
  {W.}~\bibnamefont {Alef}}, \bibinfo {author} {\bibfnamefont {K.}~\bibnamefont
  {Asada}}, \bibinfo {author} {\bibfnamefont {R.}~\bibnamefont {Azulay}},
  \bibinfo {author} {\bibfnamefont {A.-K.}\ \bibnamefont {Baczko}}, \bibinfo
  {author} {\bibfnamefont {D.}~\bibnamefont {Ball}}, \bibinfo {author}
  {\bibfnamefont {M.}~\bibnamefont {Baloković}}, \bibinfo {author}
  {\bibfnamefont {J.}~\bibnamefont {Barrett}}, \bibinfo {author} {\bibfnamefont
  {D.}~\bibnamefont {Bintley}}, \emph {et~al.},\ }\bibfield  {title} {\bibinfo
  {title} {First m87 event horizon telescope results. ii. array and
  instrumentation},\ }\href {https://doi.org/10.3847/2041-8213/ab0c96}
  {\bibfield  {journal} {\bibinfo  {journal} {The Astrophysical Journal
  Letters}\ }\textbf {\bibinfo {volume} {875}},\ \bibinfo {pages} {L2}
  (\bibinfo {year} {2019}{\natexlab{b}})}\BibitemShut {NoStop}%
\bibitem [{\citenamefont {Collaboration}\ \emph
  {et~al.}(2019{\natexlab{c}})\citenamefont {Collaboration}, \citenamefont
  {Akiyama}, \citenamefont {Alberdi}, \citenamefont {Alef}, \citenamefont
  {Asada}, \citenamefont {Azulay}, \citenamefont {Baczko}, \citenamefont
  {Ball}, \citenamefont {Baloković}, \citenamefont {Barrett}, \citenamefont
  {Bintley} \emph {et~al.}}]{3Akiyama_2019}%
  \BibitemOpen
  \bibfield  {author} {\bibinfo {author} {\bibfnamefont {T.~E. H.~T.}\
  \bibnamefont {Collaboration}}, \bibinfo {author} {\bibfnamefont
  {K.}~\bibnamefont {Akiyama}}, \bibinfo {author} {\bibfnamefont
  {A.}~\bibnamefont {Alberdi}}, \bibinfo {author} {\bibfnamefont
  {W.}~\bibnamefont {Alef}}, \bibinfo {author} {\bibfnamefont {K.}~\bibnamefont
  {Asada}}, \bibinfo {author} {\bibfnamefont {R.}~\bibnamefont {Azulay}},
  \bibinfo {author} {\bibfnamefont {A.-K.}\ \bibnamefont {Baczko}}, \bibinfo
  {author} {\bibfnamefont {D.}~\bibnamefont {Ball}}, \bibinfo {author}
  {\bibfnamefont {M.}~\bibnamefont {Baloković}}, \bibinfo {author}
  {\bibfnamefont {J.}~\bibnamefont {Barrett}}, \bibinfo {author} {\bibfnamefont
  {D.}~\bibnamefont {Bintley}}, \emph {et~al.},\ }\bibfield  {title} {\bibinfo
  {title} {First m87 event horizon telescope results. iii. data processing and
  calibration},\ }\href {https://doi.org/10.3847/2041-8213/ab0c57} {\bibfield
  {journal} {\bibinfo  {journal} {The Astrophysical Journal Letters}\ }\textbf
  {\bibinfo {volume} {875}},\ \bibinfo {pages} {L3} (\bibinfo {year}
  {2019}{\natexlab{c}})}\BibitemShut {NoStop}%
\bibitem [{\citenamefont {Collaboration}\ \emph
  {et~al.}(2019{\natexlab{d}})\citenamefont {Collaboration}, \citenamefont
  {Akiyama}, \citenamefont {Alberdi}, \citenamefont {Alef}, \citenamefont
  {Asada}, \citenamefont {Azulay}, \citenamefont {Baczko}, \citenamefont
  {Ball}, \citenamefont {Baloković}, \citenamefont {Barrett}, \citenamefont
  {Bintley} \emph {et~al.}}]{4Akiyama_2019}%
  \BibitemOpen
  \bibfield  {author} {\bibinfo {author} {\bibfnamefont {T.~E. H.~T.}\
  \bibnamefont {Collaboration}}, \bibinfo {author} {\bibfnamefont
  {K.}~\bibnamefont {Akiyama}}, \bibinfo {author} {\bibfnamefont
  {A.}~\bibnamefont {Alberdi}}, \bibinfo {author} {\bibfnamefont
  {W.}~\bibnamefont {Alef}}, \bibinfo {author} {\bibfnamefont {K.}~\bibnamefont
  {Asada}}, \bibinfo {author} {\bibfnamefont {R.}~\bibnamefont {Azulay}},
  \bibinfo {author} {\bibfnamefont {A.-K.}\ \bibnamefont {Baczko}}, \bibinfo
  {author} {\bibfnamefont {D.}~\bibnamefont {Ball}}, \bibinfo {author}
  {\bibfnamefont {M.}~\bibnamefont {Baloković}}, \bibinfo {author}
  {\bibfnamefont {J.}~\bibnamefont {Barrett}}, \bibinfo {author} {\bibfnamefont
  {D.}~\bibnamefont {Bintley}}, \emph {et~al.},\ }\bibfield  {title} {\bibinfo
  {title} {First m87 event horizon telescope results. iv. imaging the central
  supermassive black hole},\ }\href {https://doi.org/10.3847/2041-8213/ab0e85}
  {\bibfield  {journal} {\bibinfo  {journal} {The Astrophysical Journal
  Letters}\ }\textbf {\bibinfo {volume} {875}},\ \bibinfo {pages} {L4}
  (\bibinfo {year} {2019}{\natexlab{d}})}\BibitemShut {NoStop}%
\bibitem [{\citenamefont {Collaboration}\ \emph
  {et~al.}(2019{\natexlab{e}})\citenamefont {Collaboration}, \citenamefont
  {Akiyama}, \citenamefont {Alberdi}, \citenamefont {Alef}, \citenamefont
  {Asada}, \citenamefont {Azulay}, \citenamefont {Baczko}, \citenamefont
  {Ball}, \citenamefont {Baloković}, \citenamefont {Barrett}, \citenamefont
  {Bintley} \emph {et~al.}}]{5Akiyama_2019}%
  \BibitemOpen
  \bibfield  {author} {\bibinfo {author} {\bibfnamefont {T.~E. H.~T.}\
  \bibnamefont {Collaboration}}, \bibinfo {author} {\bibfnamefont
  {K.}~\bibnamefont {Akiyama}}, \bibinfo {author} {\bibfnamefont
  {A.}~\bibnamefont {Alberdi}}, \bibinfo {author} {\bibfnamefont
  {W.}~\bibnamefont {Alef}}, \bibinfo {author} {\bibfnamefont {K.}~\bibnamefont
  {Asada}}, \bibinfo {author} {\bibfnamefont {R.}~\bibnamefont {Azulay}},
  \bibinfo {author} {\bibfnamefont {A.-K.}\ \bibnamefont {Baczko}}, \bibinfo
  {author} {\bibfnamefont {D.}~\bibnamefont {Ball}}, \bibinfo {author}
  {\bibfnamefont {M.}~\bibnamefont {Baloković}}, \bibinfo {author}
  {\bibfnamefont {J.}~\bibnamefont {Barrett}}, \bibinfo {author} {\bibfnamefont
  {D.}~\bibnamefont {Bintley}}, \emph {et~al.},\ }\bibfield  {title} {\bibinfo
  {title} {First m87 event horizon telescope results. v. physical origin of the
  asymmetric ring},\ }\href {https://doi.org/10.3847/2041-8213/ab0f43}
  {\bibfield  {journal} {\bibinfo  {journal} {The Astrophysical Journal
  Letters}\ }\textbf {\bibinfo {volume} {875}},\ \bibinfo {pages} {L5}
  (\bibinfo {year} {2019}{\natexlab{e}})}\BibitemShut {NoStop}%
\bibitem [{\citenamefont {Collaboration}\ \emph
  {et~al.}(2019{\natexlab{f}})\citenamefont {Collaboration}, \citenamefont
  {Akiyama}, \citenamefont {Alberdi}, \citenamefont {Alef}, \citenamefont
  {Asada}, \citenamefont {Azulay}, \citenamefont {Baczko}, \citenamefont
  {Ball}, \citenamefont {Baloković}, \citenamefont {Barrett}, \citenamefont
  {Bintley} \emph {et~al.}}]{6Akiyama_2019}%
  \BibitemOpen
  \bibfield  {author} {\bibinfo {author} {\bibfnamefont {T.~E. H.~T.}\
  \bibnamefont {Collaboration}}, \bibinfo {author} {\bibfnamefont
  {K.}~\bibnamefont {Akiyama}}, \bibinfo {author} {\bibfnamefont
  {A.}~\bibnamefont {Alberdi}}, \bibinfo {author} {\bibfnamefont
  {W.}~\bibnamefont {Alef}}, \bibinfo {author} {\bibfnamefont {K.}~\bibnamefont
  {Asada}}, \bibinfo {author} {\bibfnamefont {R.}~\bibnamefont {Azulay}},
  \bibinfo {author} {\bibfnamefont {A.-K.}\ \bibnamefont {Baczko}}, \bibinfo
  {author} {\bibfnamefont {D.}~\bibnamefont {Ball}}, \bibinfo {author}
  {\bibfnamefont {M.}~\bibnamefont {Baloković}}, \bibinfo {author}
  {\bibfnamefont {J.}~\bibnamefont {Barrett}}, \bibinfo {author} {\bibfnamefont
  {D.}~\bibnamefont {Bintley}}, \emph {et~al.},\ }\bibfield  {title} {\bibinfo
  {title} {First m87 event horizon telescope results. vi. the shadow and mass
  of the central black hole},\ }\href
  {https://doi.org/10.3847/2041-8213/ab1141} {\bibfield  {journal} {\bibinfo
  {journal} {The Astrophysical Journal Letters}\ }\textbf {\bibinfo {volume}
  {875}},\ \bibinfo {pages} {L6} (\bibinfo {year}
  {2019}{\natexlab{f}})}\BibitemShut {NoStop}%
\bibitem [{\citenamefont {Collaboration}\ \emph
  {et~al.}(2021{\natexlab{a}})\citenamefont {Collaboration}, \citenamefont
  {Akiyama}, \citenamefont {Alberdi}, \citenamefont {Alef}, \citenamefont
  {Asada}, \citenamefont {Azulay}, \citenamefont {Baczko}, \citenamefont
  {Ball}, \citenamefont {Baloković}, \citenamefont {Barrett}, \citenamefont
  {Bintley} \emph {et~al.}}]{7Akiyama_2021}%
  \BibitemOpen
  \bibfield  {author} {\bibinfo {author} {\bibfnamefont {T.~E. H.~T.}\
  \bibnamefont {Collaboration}}, \bibinfo {author} {\bibfnamefont
  {K.}~\bibnamefont {Akiyama}}, \bibinfo {author} {\bibfnamefont
  {A.}~\bibnamefont {Alberdi}}, \bibinfo {author} {\bibfnamefont
  {W.}~\bibnamefont {Alef}}, \bibinfo {author} {\bibfnamefont {K.}~\bibnamefont
  {Asada}}, \bibinfo {author} {\bibfnamefont {R.}~\bibnamefont {Azulay}},
  \bibinfo {author} {\bibfnamefont {A.-K.}\ \bibnamefont {Baczko}}, \bibinfo
  {author} {\bibfnamefont {D.}~\bibnamefont {Ball}}, \bibinfo {author}
  {\bibfnamefont {M.}~\bibnamefont {Baloković}}, \bibinfo {author}
  {\bibfnamefont {J.}~\bibnamefont {Barrett}}, \bibinfo {author} {\bibfnamefont
  {D.}~\bibnamefont {Bintley}}, \emph {et~al.},\ }\bibfield  {title} {\bibinfo
  {title} {First m87 event horizon telescope results. vii. polarization of the
  ring},\ }\href {https://doi.org/10.3847/2041-8213/abe71d} {\bibfield
  {journal} {\bibinfo  {journal} {The Astrophysical Journal Letters}\ }\textbf
  {\bibinfo {volume} {910}},\ \bibinfo {pages} {L12} (\bibinfo {year}
  {2021}{\natexlab{a}})}\BibitemShut {NoStop}%
\bibitem [{\citenamefont {Collaboration}\ \emph
  {et~al.}(2021{\natexlab{b}})\citenamefont {Collaboration}, \citenamefont
  {Akiyama}, \citenamefont {Alberdi}, \citenamefont {Alef}, \citenamefont
  {Asada}, \citenamefont {Azulay}, \citenamefont {Baczko}, \citenamefont
  {Ball}, \citenamefont {Baloković}, \citenamefont {Barrett}, \citenamefont
  {Bintley} \emph {et~al.}}]{8Akiyama_2021}%
  \BibitemOpen
  \bibfield  {author} {\bibinfo {author} {\bibfnamefont {T.~E. H.~T.}\
  \bibnamefont {Collaboration}}, \bibinfo {author} {\bibfnamefont
  {K.}~\bibnamefont {Akiyama}}, \bibinfo {author} {\bibfnamefont
  {A.}~\bibnamefont {Alberdi}}, \bibinfo {author} {\bibfnamefont
  {W.}~\bibnamefont {Alef}}, \bibinfo {author} {\bibfnamefont {K.}~\bibnamefont
  {Asada}}, \bibinfo {author} {\bibfnamefont {R.}~\bibnamefont {Azulay}},
  \bibinfo {author} {\bibfnamefont {A.-K.}\ \bibnamefont {Baczko}}, \bibinfo
  {author} {\bibfnamefont {D.}~\bibnamefont {Ball}}, \bibinfo {author}
  {\bibfnamefont {M.}~\bibnamefont {Baloković}}, \bibinfo {author}
  {\bibfnamefont {J.}~\bibnamefont {Barrett}}, \bibinfo {author} {\bibfnamefont
  {D.}~\bibnamefont {Bintley}}, \emph {et~al.},\ }\bibfield  {title} {\bibinfo
  {title} {First m87 event horizon telescope results. viii. magnetic field
  structure near the event horizon},\ }\href
  {https://doi.org/10.3847/2041-8213/abe4de} {\bibfield  {journal} {\bibinfo
  {journal} {The Astrophysical Journal Letters}\ }\textbf {\bibinfo {volume}
  {910}},\ \bibinfo {pages} {L13} (\bibinfo {year}
  {2021}{\natexlab{b}})}\BibitemShut {NoStop}%
\bibitem [{\citenamefont {{Schwarzschild}}(1916)}]{1916skpa.conf..424S}%
  \BibitemOpen
  \bibfield  {author} {\bibinfo {author} {\bibfnamefont {K.}~\bibnamefont
  {{Schwarzschild}}},\ }\bibfield  {title} {\bibinfo {title} {{{\"U}ber das
  Gravitationsfeld einer Kugel aus inkompressibler Fl{\"u}ssigkeit nach der
  Einsteinschen Theorie}},\ }in\ \href@noop {} {\emph {\bibinfo {booktitle}
  {Sitzungsberichte der K{\"o}niglich Preussischen Akademie der Wissenschaften
  zu Berlin}}}\ (\bibinfo {year} {1916})\ pp.\ \bibinfo {pages}
  {424--434}\BibitemShut {NoStop}%
\bibitem [{\citenamefont {Ferrer}\ \emph {et~al.}(2017)\citenamefont {Ferrer},
  \citenamefont {Medeiros~da Rosa},\ and\ \citenamefont
  {Will}}]{PhysRevD.96.083014}%
  \BibitemOpen
  \bibfield  {author} {\bibinfo {author} {\bibfnamefont {F.}~\bibnamefont
  {Ferrer}}, \bibinfo {author} {\bibfnamefont {A.}~\bibnamefont {Medeiros~da
  Rosa}},\ and\ \bibinfo {author} {\bibfnamefont {C.~M.}\ \bibnamefont
  {Will}},\ }\bibfield  {title} {\bibinfo {title} {Dark matter spikes in the
  vicinity of kerr black holes},\ }\href
  {https://doi.org/10.1103/PhysRevD.96.083014} {\bibfield  {journal} {\bibinfo
  {journal} {Phys. Rev. D}\ }\textbf {\bibinfo {volume} {96}},\ \bibinfo
  {pages} {083014} (\bibinfo {year} {2017})}\BibitemShut {NoStop}%
\bibitem [{\citenamefont {Nampalliwar}\ \emph {et~al.}(2021)\citenamefont
  {Nampalliwar}, \citenamefont {Kumar}, \citenamefont {Jusufi}, \citenamefont
  {Wu}, \citenamefont {Jamil},\ and\ \citenamefont
  {Salucci}}]{Nampalliwar_2021}%
  \BibitemOpen
  \bibfield  {author} {\bibinfo {author} {\bibfnamefont {S.}~\bibnamefont
  {Nampalliwar}}, \bibinfo {author} {\bibfnamefont {S.}~\bibnamefont {Kumar}},
  \bibinfo {author} {\bibfnamefont {K.}~\bibnamefont {Jusufi}}, \bibinfo
  {author} {\bibfnamefont {Q.}~\bibnamefont {Wu}}, \bibinfo {author}
  {\bibfnamefont {M.}~\bibnamefont {Jamil}},\ and\ \bibinfo {author}
  {\bibfnamefont {P.}~\bibnamefont {Salucci}},\ }\bibfield  {title} {\bibinfo
  {title} {Modeling the sgr a* black hole immersed in a dark matter spike},\
  }\href {https://doi.org/10.3847/1538-4357/ac05cc} {\bibfield  {journal}
  {\bibinfo  {journal} {The Astrophysical Journal}\ }\textbf {\bibinfo {volume}
  {916}},\ \bibinfo {pages} {116} (\bibinfo {year} {2021})}\BibitemShut
  {NoStop}%
\bibitem [{\citenamefont {Xu}\ \emph {et~al.}(2021)\citenamefont {Xu},
  \citenamefont {Wang},\ and\ \citenamefont {Tang}}]{Xu_2021}%
  \BibitemOpen
  \bibfield  {author} {\bibinfo {author} {\bibfnamefont {Z.}~\bibnamefont
  {Xu}}, \bibinfo {author} {\bibfnamefont {J.}~\bibnamefont {Wang}},\ and\
  \bibinfo {author} {\bibfnamefont {M.}~\bibnamefont {Tang}},\ }\bibfield
  {title} {\bibinfo {title} {Deformed black hole immersed in dark matter
  spike},\ }\href {https://doi.org/10.1088/1475-7516/2021/09/007} {\bibfield
  {journal} {\bibinfo  {journal} {Journal of Cosmology and Astroparticle
  Physics}\ }\textbf {\bibinfo {volume} {2021}}\bibinfo  {number} { (09)},\
  \bibinfo {pages} {007}}\BibitemShut {NoStop}%
\bibitem [{\citenamefont {Jusufi}\ \emph {et~al.}(2020)\citenamefont {Jusufi},
  \citenamefont {Jamil},\ and\ \citenamefont {Zhu}}]{jusufi2020shadows}%
  \BibitemOpen
\bibfield  {number} {  }\bibfield  {author} {\bibinfo {author} {\bibfnamefont
  {K.}~\bibnamefont {Jusufi}}, \bibinfo {author} {\bibfnamefont
  {M.}~\bibnamefont {Jamil}},\ and\ \bibinfo {author} {\bibfnamefont
  {T.}~\bibnamefont {Zhu}},\ }\bibfield  {title} {\bibinfo {title} {Shadows of
  sgr a$*$ black hole surrounded by superfluid dark matter halo},\ }\href
  {https://doi.org/10.1140/epjc/s10052-020-7899-5} {\bibfield  {journal}
  {\bibinfo  {journal} {The European Physical Journal C}\ }\textbf {\bibinfo
  {volume} {80}},\ \bibinfo {pages} {354} (\bibinfo {year} {2020})}\BibitemShut
  {NoStop}%
\bibitem [{\citenamefont {Speeney}\ \emph {et~al.}(2022)\citenamefont
  {Speeney}, \citenamefont {Antonelli}, \citenamefont {Baibhav},\ and\
  \citenamefont {Berti}}]{PhysRevD.106.044027}%
  \BibitemOpen
  \bibfield  {author} {\bibinfo {author} {\bibfnamefont {N.}~\bibnamefont
  {Speeney}}, \bibinfo {author} {\bibfnamefont {A.}~\bibnamefont {Antonelli}},
  \bibinfo {author} {\bibfnamefont {V.}~\bibnamefont {Baibhav}},\ and\ \bibinfo
  {author} {\bibfnamefont {E.}~\bibnamefont {Berti}},\ }\bibfield  {title}
  {\bibinfo {title} {Impact of relativistic corrections on the detectability of
  dark-matter spikes with gravitational waves},\ }\href
  {https://doi.org/10.1103/PhysRevD.106.044027} {\bibfield  {journal} {\bibinfo
   {journal} {Phys. Rev. D}\ }\textbf {\bibinfo {volume} {106}},\ \bibinfo
  {pages} {044027} (\bibinfo {year} {2022})}\BibitemShut {NoStop}%
\bibitem [{\citenamefont {Lin}\ and\ \citenamefont
  {Li}(2019)}]{10.1093/mnras/stz1698}%
  \BibitemOpen
  \bibfield  {author} {\bibinfo {author} {\bibfnamefont {H.-N.}\ \bibnamefont
  {Lin}}\ and\ \bibinfo {author} {\bibfnamefont {X.}~\bibnamefont {Li}},\
  }\bibfield  {title} {\bibinfo {title} {{The dark matter profiles in the Milky
  Way}},\ }\href {https://doi.org/10.1093/mnras/stz1698} {\bibfield  {journal}
  {\bibinfo  {journal} {Monthly Notices of the Royal Astronomical Society}\
  }\textbf {\bibinfo {volume} {487}},\ \bibinfo {pages} {5679} (\bibinfo {year}
  {2019})},\ \Eprint
  {https://arxiv.org/abs/https://academic.oup.com/mnras/article-pdf/487/4/5679/28897928/stz1698\_supplemental\_file.pdf}
  {https://academic.oup.com/mnras/article-pdf/487/4/5679/28897928/stz1698\_supplemental\_file.pdf}
  \BibitemShut {NoStop}%
\bibitem [{\citenamefont {Hernquist}(1990)}]{hernquist1990analytical}%
  \BibitemOpen
  \bibfield  {author} {\bibinfo {author} {\bibfnamefont {L.}~\bibnamefont
  {Hernquist}},\ }\bibfield  {title} {\bibinfo {title} {An analytical model for
  spherical galaxies and bulges},\ }\href@noop {} {\bibfield  {journal}
  {\bibinfo  {journal} {Astrophysical Journal, Part 1 (ISSN 0004-637X), vol.
  356, June 20, 1990, p. 359-364.}\ }\textbf {\bibinfo {volume} {356}},\
  \bibinfo {pages} {359} (\bibinfo {year} {1990})}\BibitemShut {NoStop}%
\bibitem [{\citenamefont {Navarro}\ \emph {et~al.}(1997)\citenamefont
  {Navarro}, \citenamefont {Frenk},\ and\ \citenamefont
  {White}}]{Navarro_1997}%
  \BibitemOpen
  \bibfield  {author} {\bibinfo {author} {\bibfnamefont {J.~F.}\ \bibnamefont
  {Navarro}}, \bibinfo {author} {\bibfnamefont {C.~S.}\ \bibnamefont {Frenk}},\
  and\ \bibinfo {author} {\bibfnamefont {S.~D.~M.}\ \bibnamefont {White}},\
  }\bibfield  {title} {\bibinfo {title} {A universal density profile from
  hierarchical clustering},\ }\href {https://doi.org/10.1086/304888} {\bibfield
   {journal} {\bibinfo  {journal} {The Astrophysical Journal}\ }\textbf
  {\bibinfo {volume} {490}},\ \bibinfo {pages} {493} (\bibinfo {year}
  {1997})}\BibitemShut {NoStop}%
\bibitem [{\citenamefont {Gondolo}\ and\ \citenamefont
  {Silk}(1999)}]{PhysRevLett.83.1719}%
  \BibitemOpen
  \bibfield  {author} {\bibinfo {author} {\bibfnamefont {P.}~\bibnamefont
  {Gondolo}}\ and\ \bibinfo {author} {\bibfnamefont {J.}~\bibnamefont {Silk}},\
  }\bibfield  {title} {\bibinfo {title} {Dark matter annihilation at the
  galactic center},\ }\href {https://doi.org/10.1103/PhysRevLett.83.1719}
  {\bibfield  {journal} {\bibinfo  {journal} {Phys. Rev. Lett.}\ }\textbf
  {\bibinfo {volume} {83}},\ \bibinfo {pages} {1719} (\bibinfo {year}
  {1999})}\BibitemShut {NoStop}%
\bibitem [{\citenamefont {Sadeghian}\ \emph {et~al.}(2013)\citenamefont
  {Sadeghian}, \citenamefont {Ferrer},\ and\ \citenamefont
  {Will}}]{PhysRevD.88.063522}%
  \BibitemOpen
  \bibfield  {author} {\bibinfo {author} {\bibfnamefont {L.}~\bibnamefont
  {Sadeghian}}, \bibinfo {author} {\bibfnamefont {F.}~\bibnamefont {Ferrer}},\
  and\ \bibinfo {author} {\bibfnamefont {C.~M.}\ \bibnamefont {Will}},\
  }\bibfield  {title} {\bibinfo {title} {Dark-matter distributions around
  massive black holes: A general relativistic analysis},\ }\href
  {https://doi.org/10.1103/PhysRevD.88.063522} {\bibfield  {journal} {\bibinfo
  {journal} {Phys. Rev. D}\ }\textbf {\bibinfo {volume} {88}},\ \bibinfo
  {pages} {063522} (\bibinfo {year} {2013})}\BibitemShut {NoStop}%
\bibitem [{\citenamefont {Chandrasekhar}\ and\ \citenamefont
  {Thorne}(1998)}]{chandrasekhar1985mathematical}%
  \BibitemOpen
  \bibfield  {author} {\bibinfo {author} {\bibfnamefont {S.}~\bibnamefont
  {Chandrasekhar}}\ and\ \bibinfo {author} {\bibfnamefont {K.~S.}\ \bibnamefont
  {Thorne}},\ }\href@noop {} {\emph {\bibinfo {title} {The mathematical theory
  of black holes}}},\ Vol.~\bibinfo {volume} {69}\ (\bibinfo  {publisher}
  {Oxford university press},\ \bibinfo {year} {1998})\BibitemShut {NoStop}%
\bibitem [{\citenamefont {Kokkotas}\ and\ \citenamefont
  {Schmidt}(1999)}]{kokkotas1999quasi}%
  \BibitemOpen
  \bibfield  {author} {\bibinfo {author} {\bibfnamefont {K.~D.}\ \bibnamefont
  {Kokkotas}}\ and\ \bibinfo {author} {\bibfnamefont {B.~G.}\ \bibnamefont
  {Schmidt}},\ }\bibfield  {title} {\bibinfo {title} {Quasi-normal modes of
  stars and black holes},\ }\href {https://doi.org/10.12942/lrr-1999-2}
  {\bibfield  {journal} {\bibinfo  {journal} {Living Reviews in Relativity}\
  }\textbf {\bibinfo {volume} {2}},\ \bibinfo {pages} {1} (\bibinfo {year}
  {1999})}\BibitemShut {NoStop}%
\bibitem [{\citenamefont {Nollert}(1999)}]{Hans-PeterNollert_1999}%
  \BibitemOpen
  \bibfield  {author} {\bibinfo {author} {\bibfnamefont {H.-P.}\ \bibnamefont
  {Nollert}},\ }\bibfield  {title} {\bibinfo {title} {Quasinormal modes: the
  characteristic `sound' of black holes and neutron stars},\ }\href
  {https://doi.org/10.1088/0264-9381/16/12/201} {\bibfield  {journal} {\bibinfo
   {journal} {Classical and Quantum Gravity}\ }\textbf {\bibinfo {volume}
  {16}},\ \bibinfo {pages} {R159} (\bibinfo {year} {1999})}\BibitemShut
  {NoStop}%
\bibitem [{\citenamefont {Berti}\ \emph {et~al.}(2009)\citenamefont {Berti},
  \citenamefont {Cardoso},\ and\ \citenamefont {Starinets}}]{Berti_2009}%
  \BibitemOpen
  \bibfield  {author} {\bibinfo {author} {\bibfnamefont {E.}~\bibnamefont
  {Berti}}, \bibinfo {author} {\bibfnamefont {V.}~\bibnamefont {Cardoso}},\
  and\ \bibinfo {author} {\bibfnamefont {A.~O.}\ \bibnamefont {Starinets}},\
  }\bibfield  {title} {\bibinfo {title} {Quasinormal modes of black holes and
  black branes},\ }\href {https://doi.org/10.1088/0264-9381/26/16/163001}
  {\bibfield  {journal} {\bibinfo  {journal} {Classical and Quantum Gravity}\
  }\textbf {\bibinfo {volume} {26}},\ \bibinfo {pages} {163001} (\bibinfo
  {year} {2009})}\BibitemShut {NoStop}%
\bibitem [{\citenamefont {Konoplya}\ and\ \citenamefont
  {Zhidenko}(2011)}]{RevModPhys.83.793}%
  \BibitemOpen
  \bibfield  {author} {\bibinfo {author} {\bibfnamefont {R.~A.}\ \bibnamefont
  {Konoplya}}\ and\ \bibinfo {author} {\bibfnamefont {A.}~\bibnamefont
  {Zhidenko}},\ }\bibfield  {title} {\bibinfo {title} {Quasinormal modes of
  black holes: From astrophysics to string theory},\ }\href
  {https://doi.org/10.1103/RevModPhys.83.793} {\bibfield  {journal} {\bibinfo
  {journal} {Rev. Mod. Phys.}\ }\textbf {\bibinfo {volume} {83}},\ \bibinfo
  {pages} {793} (\bibinfo {year} {2011})}\BibitemShut {NoStop}%
\bibitem [{\citenamefont {{Feng}}\ \emph {et~al.}(2022)\citenamefont {{Feng}},
  \citenamefont {{Parisi}}, \citenamefont {{Chen}},\ and\ \citenamefont
  {{Lin}}}]{2022JCAP...08..032F}%
  \BibitemOpen
  \bibfield  {author} {\bibinfo {author} {\bibfnamefont {W.-X.}\ \bibnamefont
  {{Feng}}}, \bibinfo {author} {\bibfnamefont {A.}~\bibnamefont {{Parisi}}},
  \bibinfo {author} {\bibfnamefont {C.-S.}\ \bibnamefont {{Chen}}},\ and\
  \bibinfo {author} {\bibfnamefont {F.-L.}\ \bibnamefont {{Lin}}},\ }\bibfield
  {title} {\bibinfo {title} {{Self-interacting dark scalar spikes around black
  holes via relativistic Bondi accretion}},\ }\href
  {https://doi.org/10.1088/1475-7516/2022/08/032} {\bibfield  {journal}
  {\bibinfo  {journal} {Journal of Cosmology and Astroparticle Physics}\
  }\textbf {\bibinfo {volume} {2022}}\bibfield  {number} {\bibinfo  {number} {
  (8)},\ \bibinfo {eid} {032}},\ }\Eprint {https://arxiv.org/abs/2112.05160}
  {arXiv:2112.05160 [astro-ph.HE]} \BibitemShut {NoStop}%
\bibitem [{\citenamefont {Shapiro}\ and\ \citenamefont
  {Paschalidis}(2014)}]{PhysRevD.89.023506}%
  \BibitemOpen
  \bibfield  {author} {\bibinfo {author} {\bibfnamefont {S.~L.}\ \bibnamefont
  {Shapiro}}\ and\ \bibinfo {author} {\bibfnamefont {V.}~\bibnamefont
  {Paschalidis}},\ }\bibfield  {title} {\bibinfo {title} {Self-interacting dark
  matter cusps around massive black holes},\ }\href
  {https://doi.org/10.1103/PhysRevD.89.023506} {\bibfield  {journal} {\bibinfo
  {journal} {Phys. Rev. D}\ }\textbf {\bibinfo {volume} {89}},\ \bibinfo
  {pages} {023506} (\bibinfo {year} {2014})}\BibitemShut {NoStop}%
\bibitem [{\citenamefont {Luca}\ and\ \citenamefont
  {Khoury}(2023)}]{deluca2023superfluid}%
  \BibitemOpen
  \bibfield  {author} {\bibinfo {author} {\bibfnamefont {V.~D.}\ \bibnamefont
  {Luca}}\ and\ \bibinfo {author} {\bibfnamefont {J.}~\bibnamefont {Khoury}},\
  }\href@noop {} {\bibinfo {title} {Superfluid dark matter around black holes}}
  (\bibinfo {year} {2023}),\ \Eprint {https://arxiv.org/abs/2302.10286}
  {arXiv:2302.10286 [astro-ph.CO]} \BibitemShut {NoStop}%
\bibitem [{\citenamefont {Cotesta}\ \emph {et~al.}(2022)\citenamefont
  {Cotesta}, \citenamefont {Carullo}, \citenamefont {Berti},\ and\
  \citenamefont {Cardoso}}]{PhysRevLett.129.111102}%
  \BibitemOpen
  \bibfield  {author} {\bibinfo {author} {\bibfnamefont {R.}~\bibnamefont
  {Cotesta}}, \bibinfo {author} {\bibfnamefont {G.}~\bibnamefont {Carullo}},
  \bibinfo {author} {\bibfnamefont {E.}~\bibnamefont {Berti}},\ and\ \bibinfo
  {author} {\bibfnamefont {V.}~\bibnamefont {Cardoso}},\ }\bibfield  {title}
  {\bibinfo {title} {Analysis of ringdown overtones in gw150914},\ }\href
  {https://doi.org/10.1103/PhysRevLett.129.111102} {\bibfield  {journal}
  {\bibinfo  {journal} {Phys. Rev. Lett.}\ }\textbf {\bibinfo {volume} {129}},\
  \bibinfo {pages} {111102} (\bibinfo {year} {2022})}\BibitemShut {NoStop}%
\bibitem [{\citenamefont {Ruan}\ \emph {et~al.}(2020)\citenamefont {Ruan},
  \citenamefont {Guo}, \citenamefont {Cai},\ and\ \citenamefont
  {Zhang}}]{ruan2020taiji}%
  \BibitemOpen
  \bibfield  {author} {\bibinfo {author} {\bibfnamefont {W.-H.}\ \bibnamefont
  {Ruan}}, \bibinfo {author} {\bibfnamefont {Z.-K.}\ \bibnamefont {Guo}},
  \bibinfo {author} {\bibfnamefont {R.-G.}\ \bibnamefont {Cai}},\ and\ \bibinfo
  {author} {\bibfnamefont {Y.-Z.}\ \bibnamefont {Zhang}},\ }\bibfield  {title}
  {\bibinfo {title} {Taiji program: Gravitational-wave sources},\ }\href
  {https://doi.org/10.1142/S0217751X2050075X} {\bibfield  {journal} {\bibinfo
  {journal} {International Journal of Modern Physics A}\ }\textbf {\bibinfo
  {volume} {35}},\ \bibinfo {pages} {2050075} (\bibinfo {year}
  {2020})}\BibitemShut {NoStop}%
\bibitem [{\citenamefont {Lu}\ \emph {et~al.}(2019)\citenamefont {Lu},
  \citenamefont {Tan},\ and\ \citenamefont {Shao}}]{PhysRevD.100.044042}%
  \BibitemOpen
  \bibfield  {author} {\bibinfo {author} {\bibfnamefont {X.-Y.}\ \bibnamefont
  {Lu}}, \bibinfo {author} {\bibfnamefont {Y.-J.}\ \bibnamefont {Tan}},\ and\
  \bibinfo {author} {\bibfnamefont {C.-G.}\ \bibnamefont {Shao}},\ }\bibfield
  {title} {\bibinfo {title} {Sensitivity functions for space-borne
  gravitational wave detectors},\ }\href
  {https://doi.org/10.1103/PhysRevD.100.044042} {\bibfield  {journal} {\bibinfo
   {journal} {Phys. Rev. D}\ }\textbf {\bibinfo {volume} {100}},\ \bibinfo
  {pages} {044042} (\bibinfo {year} {2019})}\BibitemShut {NoStop}%
\bibitem [{\citenamefont {Moore}\ \emph {et~al.}(2014)\citenamefont {Moore},
  \citenamefont {Cole},\ and\ \citenamefont {Berry}}]{Moore_2015}%
  \BibitemOpen
  \bibfield  {author} {\bibinfo {author} {\bibfnamefont {C.~J.}\ \bibnamefont
  {Moore}}, \bibinfo {author} {\bibfnamefont {R.~H.}\ \bibnamefont {Cole}},\
  and\ \bibinfo {author} {\bibfnamefont {C.~P.~L.}\ \bibnamefont {Berry}},\
  }\bibfield  {title} {\bibinfo {title} {Gravitational-wave sensitivity
  curves},\ }\href {https://doi.org/10.1088/0264-9381/32/1/015014} {\bibfield
  {journal} {\bibinfo  {journal} {Classical and Quantum Gravity}\ }\textbf
  {\bibinfo {volume} {32}},\ \bibinfo {pages} {015014} (\bibinfo {year}
  {2014})}\BibitemShut {NoStop}%
\bibitem [{\citenamefont {Berti}\ \emph {et~al.}(2006)\citenamefont {Berti},
  \citenamefont {Cardoso},\ and\ \citenamefont {Will}}]{PhysRevD.73.064030}%
  \BibitemOpen
  \bibfield  {author} {\bibinfo {author} {\bibfnamefont {E.}~\bibnamefont
  {Berti}}, \bibinfo {author} {\bibfnamefont {V.}~\bibnamefont {Cardoso}},\
  and\ \bibinfo {author} {\bibfnamefont {C.~M.}\ \bibnamefont {Will}},\
  }\bibfield  {title} {\bibinfo {title} {Gravitational-wave spectroscopy of
  massive black holes with the space interferometer lisa},\ }\href
  {https://doi.org/10.1103/PhysRevD.73.064030} {\bibfield  {journal} {\bibinfo
  {journal} {Phys. Rev. D}\ }\textbf {\bibinfo {volume} {73}},\ \bibinfo
  {pages} {064030} (\bibinfo {year} {2006})}\BibitemShut {NoStop}%
\bibitem [{\citenamefont {Liu}\ \emph {et~al.}(2021)\citenamefont {Liu},
  \citenamefont {Yang}, \citenamefont {Wu}, \citenamefont {Xing}, \citenamefont
  {Xu},\ and\ \citenamefont {Long}}]{PhysRevD.104.104042}%
  \BibitemOpen
  \bibfield  {author} {\bibinfo {author} {\bibfnamefont {D.}~\bibnamefont
  {Liu}}, \bibinfo {author} {\bibfnamefont {Y.}~\bibnamefont {Yang}}, \bibinfo
  {author} {\bibfnamefont {S.}~\bibnamefont {Wu}}, \bibinfo {author}
  {\bibfnamefont {Y.}~\bibnamefont {Xing}}, \bibinfo {author} {\bibfnamefont
  {Z.}~\bibnamefont {Xu}},\ and\ \bibinfo {author} {\bibfnamefont {Z.-W.}\
  \bibnamefont {Long}},\ }\bibfield  {title} {\bibinfo {title} {Ringing of a
  black hole in a dark matter halo},\ }\href
  {https://doi.org/10.1103/PhysRevD.104.104042} {\bibfield  {journal} {\bibinfo
   {journal} {Phys. Rev. D}\ }\textbf {\bibinfo {volume} {104}},\ \bibinfo
  {pages} {104042} (\bibinfo {year} {2021})}\BibitemShut {NoStop}%
\bibitem [{\citenamefont {Zhang}\ \emph {et~al.}(2022)\citenamefont {Zhang},
  \citenamefont {Zhu}, \citenamefont {Fang},\ and\ \citenamefont
  {Wang}}]{ZHANG2022101078}%
  \BibitemOpen
  \bibfield  {author} {\bibinfo {author} {\bibfnamefont {C.}~\bibnamefont
  {Zhang}}, \bibinfo {author} {\bibfnamefont {T.}~\bibnamefont {Zhu}}, \bibinfo
  {author} {\bibfnamefont {X.}~\bibnamefont {Fang}},\ and\ \bibinfo {author}
  {\bibfnamefont {A.}~\bibnamefont {Wang}},\ }\bibfield  {title} {\bibinfo
  {title} {Imprints of dark matter on gravitational ringing of supermassive
  black holes},\ }\href
  {https://doi.org/https://doi.org/10.1016/j.dark.2022.101078} {\bibfield
  {journal} {\bibinfo  {journal} {Physics of the Dark Universe}\ }\textbf
  {\bibinfo {volume} {37}},\ \bibinfo {pages} {101078} (\bibinfo {year}
  {2022})}\BibitemShut {NoStop}%
\bibitem [{\citenamefont {Zhang}\ \emph {et~al.}(2021)\citenamefont {Zhang},
  \citenamefont {Zhu},\ and\ \citenamefont {Wang}}]{PhysRevD.104.124082}%
  \BibitemOpen
  \bibfield  {author} {\bibinfo {author} {\bibfnamefont {C.}~\bibnamefont
  {Zhang}}, \bibinfo {author} {\bibfnamefont {T.}~\bibnamefont {Zhu}},\ and\
  \bibinfo {author} {\bibfnamefont {A.}~\bibnamefont {Wang}},\ }\bibfield
  {title} {\bibinfo {title} {Gravitational axial perturbations of
  schwarzschild-like black holes in dark matter halos},\ }\href
  {https://doi.org/10.1103/PhysRevD.104.124082} {\bibfield  {journal} {\bibinfo
   {journal} {Phys. Rev. D}\ }\textbf {\bibinfo {volume} {104}},\ \bibinfo
  {pages} {124082} (\bibinfo {year} {2021})}\BibitemShut {NoStop}%
\bibitem [{\citenamefont {Cardoso}\ \emph
  {et~al.}(2022{\natexlab{a}})\citenamefont {Cardoso}, \citenamefont
  {Destounis}, \citenamefont {Duque}, \citenamefont {Macedo},\ and\
  \citenamefont {Maselli}}]{PhysRevD.105.L061501}%
  \BibitemOpen
  \bibfield  {author} {\bibinfo {author} {\bibfnamefont {V.}~\bibnamefont
  {Cardoso}}, \bibinfo {author} {\bibfnamefont {K.}~\bibnamefont {Destounis}},
  \bibinfo {author} {\bibfnamefont {F.}~\bibnamefont {Duque}}, \bibinfo
  {author} {\bibfnamefont {R.~P.}\ \bibnamefont {Macedo}},\ and\ \bibinfo
  {author} {\bibfnamefont {A.}~\bibnamefont {Maselli}},\ }\bibfield  {title}
  {\bibinfo {title} {Black holes in galaxies: Environmental impact on
  gravitational-wave generation and propagation},\ }\href
  {https://doi.org/10.1103/PhysRevD.105.L061501} {\bibfield  {journal}
  {\bibinfo  {journal} {Phys. Rev. D}\ }\textbf {\bibinfo {volume} {105}},\
  \bibinfo {pages} {L061501} (\bibinfo {year}
  {2022}{\natexlab{a}})}\BibitemShut {NoStop}%
\bibitem [{\citenamefont {Konoplya}(2021)}]{KONOPLYA2021136734}%
  \BibitemOpen
  \bibfield  {author} {\bibinfo {author} {\bibfnamefont {R.}~\bibnamefont
  {Konoplya}},\ }\bibfield  {title} {\bibinfo {title} {Black holes in galactic
  centers: Quasinormal ringing, grey-body factors and unruh temperature},\
  }\href {https://doi.org/https://doi.org/10.1016/j.physletb.2021.136734}
  {\bibfield  {journal} {\bibinfo  {journal} {Physics Letters B}\ }\textbf
  {\bibinfo {volume} {823}},\ \bibinfo {pages} {136734} (\bibinfo {year}
  {2021})}\BibitemShut {NoStop}%
\bibitem [{\citenamefont {Daghigh}\ and\ \citenamefont
  {Kunstatter}(2022)}]{Daghigh_2022}%
  \BibitemOpen
  \bibfield  {author} {\bibinfo {author} {\bibfnamefont {R.~G.}\ \bibnamefont
  {Daghigh}}\ and\ \bibinfo {author} {\bibfnamefont {G.}~\bibnamefont
  {Kunstatter}},\ }\bibfield  {title} {\bibinfo {title} {Spacetime metrics and
  ringdown waveforms for galactic black holes surrounded by a dark matter
  spike},\ }\href {https://doi.org/10.3847/1538-4357/ac940b} {\bibfield
  {journal} {\bibinfo  {journal} {The Astrophysical Journal}\ }\textbf
  {\bibinfo {volume} {940}},\ \bibinfo {pages} {33} (\bibinfo {year}
  {2022})}\BibitemShut {NoStop}%
\bibitem [{\citenamefont {Konoplya}\ and\ \citenamefont
  {Zhidenko}(2022)}]{Konoplya_2022}%
  \BibitemOpen
  \bibfield  {author} {\bibinfo {author} {\bibfnamefont {R.~A.}\ \bibnamefont
  {Konoplya}}\ and\ \bibinfo {author} {\bibfnamefont {A.}~\bibnamefont
  {Zhidenko}},\ }\bibfield  {title} {\bibinfo {title} {Solutions of the
  einstein equations for a black hole surrounded by a galactic halo},\ }\href
  {https://doi.org/10.3847/1538-4357/ac76bc} {\bibfield  {journal} {\bibinfo
  {journal} {The Astrophysical Journal}\ }\textbf {\bibinfo {volume} {933}},\
  \bibinfo {pages} {166} (\bibinfo {year} {2022})}\BibitemShut {NoStop}%
\bibitem [{\citenamefont {Carroll}(2019)}]{carroll2019spacetime}%
  \BibitemOpen
  \bibfield  {author} {\bibinfo {author} {\bibfnamefont {S.~M.}\ \bibnamefont
  {Carroll}},\ }\href@noop {} {\emph {\bibinfo {title} {Spacetime and
  geometry}}}\ (\bibinfo  {publisher} {Cambridge University Press},\ \bibinfo
  {year} {2019})\BibitemShut {NoStop}%
\bibitem [{\citenamefont {Abramowitz}\ and\ \citenamefont
  {Stegun}(1970)}]{abramowitz3handbook}%
  \BibitemOpen
  \bibfield  {author} {\bibinfo {author} {\bibfnamefont {M.}~\bibnamefont
  {Abramowitz}}\ and\ \bibinfo {author} {\bibfnamefont {I.}~\bibnamefont
  {Stegun}},\ }\href@noop {} {\bibinfo {title} {Handbook of mathematical
  functions (dover publications inc., new york).}} (\bibinfo {year}
  {1970})\BibitemShut {NoStop}%
\bibitem [{\citenamefont {Olver}\ \emph {et~al.}(2010)\citenamefont {Olver},
  \citenamefont {Lozier}, \citenamefont {Boisvert},\ and\ \citenamefont
  {Clark}}]{olver2010nist}%
  \BibitemOpen
  \bibfield  {author} {\bibinfo {author} {\bibfnamefont {F.~W.}\ \bibnamefont
  {Olver}}, \bibinfo {author} {\bibfnamefont {D.~W.}\ \bibnamefont {Lozier}},
  \bibinfo {author} {\bibfnamefont {R.~F.}\ \bibnamefont {Boisvert}},\ and\
  \bibinfo {author} {\bibfnamefont {C.~W.}\ \bibnamefont {Clark}},\ }\href@noop
  {} {\emph {\bibinfo {title} {NIST handbook of mathematical functions hardback
  and CD-ROM}}}\ (\bibinfo  {publisher} {Cambridge university press},\ \bibinfo
  {year} {2010})\BibitemShut {NoStop}%
\bibitem [{\citenamefont {{Futterman}}\ \emph {et~al.}(1988)\citenamefont
  {{Futterman}}, \citenamefont {{Handler}},\ and\ \citenamefont
  {{Matzner}}}]{1988sfbh.book.....F}%
  \BibitemOpen
  \bibfield  {author} {\bibinfo {author} {\bibfnamefont {J.~A.~H.}\
  \bibnamefont {{Futterman}}}, \bibinfo {author} {\bibfnamefont {F.~A.}\
  \bibnamefont {{Handler}}},\ and\ \bibinfo {author} {\bibfnamefont {R.~A.}\
  \bibnamefont {{Matzner}}},\ }\href
  {https://ui.adsabs.harvard.edu/abs/1988sfbh.book.....F} {\emph {\bibinfo
  {title} {{Scattering from black holes}}}}\ (\bibinfo  {publisher} {Cambridge
  ; New York : Cambridge University Press},\ \bibinfo {year}
  {1988})\BibitemShut {NoStop}%
\bibitem [{\citenamefont {Davis}\ \emph {et~al.}(1972)\citenamefont {Davis},
  \citenamefont {Ruffini},\ and\ \citenamefont {Tiomno}}]{SchFallingParticle}%
  \BibitemOpen
  \bibfield  {author} {\bibinfo {author} {\bibfnamefont {M.}~\bibnamefont
  {Davis}}, \bibinfo {author} {\bibfnamefont {R.}~\bibnamefont {Ruffini}},\
  and\ \bibinfo {author} {\bibfnamefont {J.}~\bibnamefont {Tiomno}},\
  }\bibfield  {title} {\bibinfo {title} {Pulses of gravitational radiation of a
  particle falling radially into a schwarzschild black hole},\ }\href
  {https://doi.org/10.1103/PhysRevD.5.2932} {\bibfield  {journal} {\bibinfo
  {journal} {Phys. Rev. D}\ }\textbf {\bibinfo {volume} {5}},\ \bibinfo {pages}
  {2932} (\bibinfo {year} {1972})}\BibitemShut {NoStop}%
\bibitem [{\citenamefont {Thompson}\ \emph {et~al.}(2017)\citenamefont
  {Thompson}, \citenamefont {Chen},\ and\ \citenamefont
  {Whiting}}]{Thompson_2017}%
  \BibitemOpen
  \bibfield  {author} {\bibinfo {author} {\bibfnamefont {J.~E.}\ \bibnamefont
  {Thompson}}, \bibinfo {author} {\bibfnamefont {H.}~\bibnamefont {Chen}},\
  and\ \bibinfo {author} {\bibfnamefont {B.~F.}\ \bibnamefont {Whiting}},\
  }\bibfield  {title} {\bibinfo {title} {Gauge invariant perturbations of the
  schwarzschild spacetime},\ }\href {https://doi.org/10.1088/1361-6382/aa7f5b}
  {\bibfield  {journal} {\bibinfo  {journal} {Classical and Quantum Gravity}\
  }\textbf {\bibinfo {volume} {34}},\ \bibinfo {pages} {174001} (\bibinfo
  {year} {2017})}\BibitemShut {NoStop}%
\bibitem [{\citenamefont {Nagar}\ and\ \citenamefont
  {Rezzolla}(2005)}]{Nagar_2005}%
  \BibitemOpen
  \bibfield  {author} {\bibinfo {author} {\bibfnamefont {A.}~\bibnamefont
  {Nagar}}\ and\ \bibinfo {author} {\bibfnamefont {L.}~\bibnamefont
  {Rezzolla}},\ }\bibfield  {title} {\bibinfo {title} {Gauge-invariant
  non-spherical metric perturbations of schwarzschild black-hole spacetimes},\
  }\href {https://doi.org/10.1088/0264-9381/22/16/R01} {\bibfield  {journal}
  {\bibinfo  {journal} {Classical and Quantum Gravity}\ }\textbf {\bibinfo
  {volume} {22}},\ \bibinfo {pages} {R167} (\bibinfo {year}
  {2005})}\BibitemShut {NoStop}%
\bibitem [{\citenamefont {Vishveshwara}(1970)}]{PhysRevD.1.2870}%
  \BibitemOpen
  \bibfield  {author} {\bibinfo {author} {\bibfnamefont {C.~V.}\ \bibnamefont
  {Vishveshwara}},\ }\bibfield  {title} {\bibinfo {title} {Stability of the
  schwarzschild metric},\ }\href {https://doi.org/10.1103/PhysRevD.1.2870}
  {\bibfield  {journal} {\bibinfo  {journal} {Phys. Rev. D}\ }\textbf {\bibinfo
  {volume} {1}},\ \bibinfo {pages} {2870} (\bibinfo {year} {1970})}\BibitemShut
  {NoStop}%
\bibitem [{\citenamefont {Cardoso}\ \emph
  {et~al.}(2022{\natexlab{b}})\citenamefont {Cardoso}, \citenamefont
  {Destounis}, \citenamefont {Duque}, \citenamefont {Macedo},\ and\
  \citenamefont {Maselli}}]{PhysRevLett.129.241103}%
  \BibitemOpen
  \bibfield  {author} {\bibinfo {author} {\bibfnamefont {V.}~\bibnamefont
  {Cardoso}}, \bibinfo {author} {\bibfnamefont {K.}~\bibnamefont {Destounis}},
  \bibinfo {author} {\bibfnamefont {F.}~\bibnamefont {Duque}}, \bibinfo
  {author} {\bibfnamefont {R.~P.}\ \bibnamefont {Macedo}},\ and\ \bibinfo
  {author} {\bibfnamefont {A.}~\bibnamefont {Maselli}},\ }\bibfield  {title}
  {\bibinfo {title} {Gravitational waves from extreme-mass-ratio systems in
  astrophysical environments},\ }\href
  {https://doi.org/10.1103/PhysRevLett.129.241103} {\bibfield  {journal}
  {\bibinfo  {journal} {Phys. Rev. Lett.}\ }\textbf {\bibinfo {volume} {129}},\
  \bibinfo {pages} {241103} (\bibinfo {year} {2022}{\natexlab{b}})}\BibitemShut
  {NoStop}%
\bibitem [{\citenamefont {Liu}\ \emph {et~al.}(2023)\citenamefont {Liu},
  \citenamefont {Fang}, \citenamefont {Jing},\ and\ \citenamefont
  {Wang}}]{liu2023gauge}%
  \BibitemOpen
  \bibfield  {author} {\bibinfo {author} {\bibfnamefont {W.}~\bibnamefont
  {Liu}}, \bibinfo {author} {\bibfnamefont {X.}~\bibnamefont {Fang}}, \bibinfo
  {author} {\bibfnamefont {J.}~\bibnamefont {Jing}},\ and\ \bibinfo {author}
  {\bibfnamefont {A.}~\bibnamefont {Wang}},\ }\bibfield  {title} {\bibinfo
  {title} {Gauge invariant perturbations of general spherically symmetric
  spacetimes},\ }\href {https://doi.org/10.1007/s11433-022-1956-4} {\bibfield
  {journal} {\bibinfo  {journal} {Science China Physics, Mechanics \&
  Astronomy}\ }\textbf {\bibinfo {volume} {66}},\ \bibinfo {pages} {210411}
  (\bibinfo {year} {2023})}\BibitemShut {NoStop}%
\bibitem [{\citenamefont {Regge}\ and\ \citenamefont {Wheeler}(1957)}]{RW}%
  \BibitemOpen
  \bibfield  {author} {\bibinfo {author} {\bibfnamefont {T.}~\bibnamefont
  {Regge}}\ and\ \bibinfo {author} {\bibfnamefont {J.~A.}\ \bibnamefont
  {Wheeler}},\ }\bibfield  {title} {\bibinfo {title} {Stability of a
  schwarzschild singularity},\ }\href
  {https://doi.org/10.1103/PhysRev.108.1063} {\bibfield  {journal} {\bibinfo
  {journal} {Phys. Rev.}\ }\textbf {\bibinfo {volume} {108}},\ \bibinfo {pages}
  {1063} (\bibinfo {year} {1957})}\BibitemShut {NoStop}%
\bibitem [{\citenamefont {Schutz}\ and\ \citenamefont
  {Will}(1985)}]{schutz1985black}%
  \BibitemOpen
  \bibfield  {author} {\bibinfo {author} {\bibfnamefont {B.~F.}\ \bibnamefont
  {Schutz}}\ and\ \bibinfo {author} {\bibfnamefont {C.~M.}\ \bibnamefont
  {Will}},\ }\bibfield  {title} {\bibinfo {title} {Black hole normal modes: a
  semianalytic approach},\ }\href {https://doi.org/10.1086/184453} {\bibfield
  {journal} {\bibinfo  {journal} {The Astrophysical Journal}\ }\textbf
  {\bibinfo {volume} {291}},\ \bibinfo {pages} {L33} (\bibinfo {year}
  {1985})}\BibitemShut {NoStop}%
\bibitem [{\citenamefont {Iyer}\ and\ \citenamefont
  {Will}(1987)}]{PhysRevD.35.3621}%
  \BibitemOpen
  \bibfield  {author} {\bibinfo {author} {\bibfnamefont {S.}~\bibnamefont
  {Iyer}}\ and\ \bibinfo {author} {\bibfnamefont {C.~M.}\ \bibnamefont
  {Will}},\ }\bibfield  {title} {\bibinfo {title} {Black-hole normal modes: A
  wkb approach. i. foundations and application of a higher-order wkb analysis
  of potential-barrier scattering},\ }\href
  {https://doi.org/10.1103/PhysRevD.35.3621} {\bibfield  {journal} {\bibinfo
  {journal} {Phys. Rev. D}\ }\textbf {\bibinfo {volume} {35}},\ \bibinfo
  {pages} {3621} (\bibinfo {year} {1987})}\BibitemShut {NoStop}%
\bibitem [{\citenamefont {Konoplya}(2003)}]{PhysRevD.68.024018}%
  \BibitemOpen
  \bibfield  {author} {\bibinfo {author} {\bibfnamefont {R.~A.}\ \bibnamefont
  {Konoplya}},\ }\bibfield  {title} {\bibinfo {title} {Quasinormal behavior of
  the $d$-dimensional schwarzschild black hole and the higher order wkb
  approach},\ }\href {https://doi.org/10.1103/PhysRevD.68.024018} {\bibfield
  {journal} {\bibinfo  {journal} {Phys. Rev. D}\ }\textbf {\bibinfo {volume}
  {68}},\ \bibinfo {pages} {024018} (\bibinfo {year} {2003})}\BibitemShut
  {NoStop}%
\bibitem [{\citenamefont {Lin}\ and\ \citenamefont {Qian}(2017)}]{Lin_2017}%
  \BibitemOpen
  \bibfield  {author} {\bibinfo {author} {\bibfnamefont {K.}~\bibnamefont
  {Lin}}\ and\ \bibinfo {author} {\bibfnamefont {W.-L.}\ \bibnamefont {Qian}},\
  }\bibfield  {title} {\bibinfo {title} {A matrix method for quasinormal modes:
  Schwarzschild black holes in asymptotically flat and (anti-) de sitter
  spacetimes},\ }\href {https://doi.org/10.1088/1361-6382/aa6643} {\bibfield
  {journal} {\bibinfo  {journal} {Classical and Quantum Gravity}\ }\textbf
  {\bibinfo {volume} {34}},\ \bibinfo {pages} {095004} (\bibinfo {year}
  {2017})}\BibitemShut {NoStop}%
\bibitem [{\citenamefont {Lin}\ \emph {et~al.}(2017)\citenamefont {Lin},
  \citenamefont {Qian}, \citenamefont {Pavan},\ and\ \citenamefont
  {Abdalla}}]{lin2017matrix}%
  \BibitemOpen
  \bibfield  {author} {\bibinfo {author} {\bibfnamefont {K.}~\bibnamefont
  {Lin}}, \bibinfo {author} {\bibfnamefont {W.-L.}\ \bibnamefont {Qian}},
  \bibinfo {author} {\bibfnamefont {A.~B.}\ \bibnamefont {Pavan}},\ and\
  \bibinfo {author} {\bibfnamefont {E.}~\bibnamefont {Abdalla}},\ }\bibfield
  {title} {\bibinfo {title} {A matrix method for quasinormal modes: Kerr and
  kerr--sen black holes},\ }\href {https://doi.org/10.1142/S0217732317501346}
  {\bibfield  {journal} {\bibinfo  {journal} {Modern Physics Letters A}\
  }\textbf {\bibinfo {volume} {32}},\ \bibinfo {pages} {1750134} (\bibinfo
  {year} {2017})}\BibitemShut {NoStop}%
\bibitem [{\citenamefont {Lin}\ and\ \citenamefont
  {Qian}(2022)}]{https://doi.org/10.48550/arxiv.2209.11612}%
  \BibitemOpen
  \bibfield  {author} {\bibinfo {author} {\bibfnamefont {K.}~\bibnamefont
  {Lin}}\ and\ \bibinfo {author} {\bibfnamefont {W.-L.}\ \bibnamefont {Qian}},\
  }\href {https://doi.org/10.48550/ARXIV.2209.11612} {\bibinfo {title}
  {High-order matrix method with delimited expansion domain}} (\bibinfo {year}
  {2022})\BibitemShut {NoStop}%
\bibitem [{\citenamefont {Shen}\ \emph {et~al.}(2022)\citenamefont {Shen},
  \citenamefont {Qian}, \citenamefont {Lin}, \citenamefont {Shao},\ and\
  \citenamefont {Pan}}]{Shen_2022}%
  \BibitemOpen
  \bibfield  {author} {\bibinfo {author} {\bibfnamefont {S.-F.}\ \bibnamefont
  {Shen}}, \bibinfo {author} {\bibfnamefont {W.-L.}\ \bibnamefont {Qian}},
  \bibinfo {author} {\bibfnamefont {K.}~\bibnamefont {Lin}}, \bibinfo {author}
  {\bibfnamefont {C.-G.}\ \bibnamefont {Shao}},\ and\ \bibinfo {author}
  {\bibfnamefont {Y.}~\bibnamefont {Pan}},\ }\bibfield  {title} {\bibinfo
  {title} {Matrix method for perturbed black hole metric with discontinuity},\
  }\href {https://doi.org/10.1088/1361-6382/ac95f1} {\bibfield  {journal}
  {\bibinfo  {journal} {Classical and Quantum Gravity}\ }\textbf {\bibinfo
  {volume} {39}},\ \bibinfo {pages} {225004} (\bibinfo {year}
  {2022})}\BibitemShut {NoStop}%
\bibitem [{\citenamefont {Lin}(2023)}]{PhysRevD.107.124002}%
  \BibitemOpen
  \bibfield  {author} {\bibinfo {author} {\bibfnamefont {K.}~\bibnamefont
  {Lin}},\ }\bibfield  {title} {\bibinfo {title} {Quasinormal modes and echo
  effect of a cylindrical anti--de sitter black hole spacetime with a thin
  shell},\ }\href {https://doi.org/10.1103/PhysRevD.107.124002} {\bibfield
  {journal} {\bibinfo  {journal} {Phys. Rev. D}\ }\textbf {\bibinfo {volume}
  {107}},\ \bibinfo {pages} {124002} (\bibinfo {year} {2023})}\BibitemShut
  {NoStop}%
\bibitem [{\citenamefont {Zhao}\ \emph {et~al.}(2022)\citenamefont {Zhao},
  \citenamefont {Sun}, \citenamefont {Mai},\ and\ \citenamefont
  {Cao}}]{https://doi.org/10.48550/arxiv.2212.00747}%
  \BibitemOpen
  \bibfield  {author} {\bibinfo {author} {\bibfnamefont {Y.}~\bibnamefont
  {Zhao}}, \bibinfo {author} {\bibfnamefont {B.}~\bibnamefont {Sun}}, \bibinfo
  {author} {\bibfnamefont {Z.-F.}\ \bibnamefont {Mai}},\ and\ \bibinfo {author}
  {\bibfnamefont {Z.}~\bibnamefont {Cao}},\ }\href
  {https://doi.org/10.48550/ARXIV.2212.00747} {\bibinfo {title} {Quasi normal
  modes of black holes and detection in ringdown process}} (\bibinfo {year}
  {2022})\BibitemShut {NoStop}%
\bibitem [{\citenamefont {Israel}(1966)}]{israel1966nuovo}%
  \BibitemOpen
  \bibfield  {author} {\bibinfo {author} {\bibfnamefont {W.}~\bibnamefont
  {Israel}},\ }\bibfield  {title} {\bibinfo {title} {Nuovo cim b44s10 1},\
  }\href@noop {} {\bibfield  {journal} {\bibinfo  {journal} {Erratum: ibid
  Nuovo Cim B}\ }\textbf {\bibinfo {volume} {48}},\ \bibinfo {pages} {463}
  (\bibinfo {year} {1966})}\BibitemShut {NoStop}%
\bibitem [{\citenamefont {Leung}\ \emph {et~al.}(1999)\citenamefont {Leung},
  \citenamefont {Liu}, \citenamefont {Suen}, \citenamefont {Tam},\ and\
  \citenamefont {Young}}]{PhysRevD.59.044034}%
  \BibitemOpen
  \bibfield  {author} {\bibinfo {author} {\bibfnamefont {P.~T.}\ \bibnamefont
  {Leung}}, \bibinfo {author} {\bibfnamefont {Y.~T.}\ \bibnamefont {Liu}},
  \bibinfo {author} {\bibfnamefont {W.~M.}\ \bibnamefont {Suen}}, \bibinfo
  {author} {\bibfnamefont {C.~Y.}\ \bibnamefont {Tam}},\ and\ \bibinfo {author}
  {\bibfnamefont {K.}~\bibnamefont {Young}},\ }\bibfield  {title} {\bibinfo
  {title} {Perturbative approach to the quasinormal modes of dirty black
  holes},\ }\href {https://doi.org/10.1103/PhysRevD.59.044034} {\bibfield
  {journal} {\bibinfo  {journal} {Phys. Rev. D}\ }\textbf {\bibinfo {volume}
  {59}},\ \bibinfo {pages} {044034} (\bibinfo {year} {1999})}\BibitemShut
  {NoStop}%
\bibitem [{\citenamefont {Ferrari}\ and\ \citenamefont
  {Mashhoon}(1984)}]{PhysRevLett.52.1361}%
  \BibitemOpen
  \bibfield  {author} {\bibinfo {author} {\bibfnamefont {V.}~\bibnamefont
  {Ferrari}}\ and\ \bibinfo {author} {\bibfnamefont {B.}~\bibnamefont
  {Mashhoon}},\ }\bibfield  {title} {\bibinfo {title} {Oscillations of a black
  hole},\ }\href {https://doi.org/10.1103/PhysRevLett.52.1361} {\bibfield
  {journal} {\bibinfo  {journal} {Phys. Rev. Lett.}\ }\textbf {\bibinfo
  {volume} {52}},\ \bibinfo {pages} {1361} (\bibinfo {year}
  {1984})}\BibitemShut {NoStop}%
\bibitem [{\citenamefont {Shi}\ \emph {et~al.}(2019)\citenamefont {Shi},
  \citenamefont {Bao}, \citenamefont {Wang}, \citenamefont {Zhang},
  \citenamefont {Hu}, \citenamefont {Sesana}, \citenamefont {Barausse},
  \citenamefont {Mei},\ and\ \citenamefont {Luo}}]{PhysRevD.100.044036}%
  \BibitemOpen
  \bibfield  {author} {\bibinfo {author} {\bibfnamefont {C.}~\bibnamefont
  {Shi}}, \bibinfo {author} {\bibfnamefont {J.}~\bibnamefont {Bao}}, \bibinfo
  {author} {\bibfnamefont {H.-T.}\ \bibnamefont {Wang}}, \bibinfo {author}
  {\bibfnamefont {J.-d.}\ \bibnamefont {Zhang}}, \bibinfo {author}
  {\bibfnamefont {Y.-M.}\ \bibnamefont {Hu}}, \bibinfo {author} {\bibfnamefont
  {A.}~\bibnamefont {Sesana}}, \bibinfo {author} {\bibfnamefont
  {E.}~\bibnamefont {Barausse}}, \bibinfo {author} {\bibfnamefont
  {J.}~\bibnamefont {Mei}},\ and\ \bibinfo {author} {\bibfnamefont
  {J.}~\bibnamefont {Luo}},\ }\bibfield  {title} {\bibinfo {title} {Science
  with the tianqin observatory: Preliminary results on testing the no-hair
  theorem with ringdown signals},\ }\href
  {https://doi.org/10.1103/PhysRevD.100.044036} {\bibfield  {journal} {\bibinfo
   {journal} {Phys. Rev. D}\ }\textbf {\bibinfo {volume} {100}},\ \bibinfo
  {pages} {044036} (\bibinfo {year} {2019})}\BibitemShut {NoStop}%
\end{thebibliography}%

\end{document}